\documentclass[aps,prb,twocolumn,showpacs,superscriptaddress,groupedaddress]{revtex4-1}  
\usepackage{graphicx}  
\usepackage{dcolumn}   
\usepackage{bm}        
\usepackage{amssymb}   
\usepackage{mathcomp}  
\usepackage{amsmath}   
\usepackage{color}     
\usepackage{pdfpages}
\usepackage{pgffor}

\makeatletter
\AtBeginDocument{\let\LS@rot\@undefined}
\makeatother

\begin{document}

\title{Measuring anisotropic spin relaxation in graphene}     
\author{Sebastian Ringer} \affiliation{Institute of Experimental and Applied Physics, University of Regensburg, Germany}
\author{Stefan Hartl} \affiliation{Institute of Experimental and Applied Physics, University of Regensburg, Germany}
\author{Matthias Rosenauer} \affiliation{Institute of Experimental and Applied Physics, University of Regensburg, Germany}
\author{Tobias V\"olkl} \affiliation{Institute of Experimental and Applied Physics, University of Regensburg, Germany}
\author{Maximilian Kadur} \affiliation{Institute of Experimental and Applied Physics, University of Regensburg, Germany}
\author{Franz Hopperdietzel} \affiliation{Institute of Experimental and Applied Physics, University of Regensburg, Germany}
\author{Dieter Weiss} \affiliation{Institute of Experimental and Applied Physics, University of Regensburg, Germany}
\author{Jonathan Eroms} \affiliation{Institute of Experimental and Applied Physics, University of Regensburg, Germany}
\email{jonathan.eroms@ur.de}
\date{\today}

\begin{abstract}
We compare different methods to measure the anisotropy of the spin-lifetime in graphene. In addition to out-of-plane rotation of the ferromagnetic electrodes and oblique spin precession, we present a Hanle experiment where the electron spins precess around either a magnetic field perpendicular to the graphene plane or around an in-plane field. In the latter case, electrons are subject to both in-plane and out-of-plane spin relaxation. To fit the data, we use a numerical simulation that can calculate precession with anisotropies in the spin-lifetimes under magnetic fields in any direction. Our data show a small, but distinct anisotropy that can be explained by the combined action of isotropic mechanisms, such as relaxation by the contacts and resonant scattering by magnetic impurities, and an anisotropic Rashba spin-orbit based mechanism. We also assess potential sources of error in all three types of experiment and conclude that the in-plane/out-of-plane Hanle method is most reliable.

\end{abstract}

\maketitle

\section{\label{sec:introduction}Introduction}

Graphene has been proposed as a promising material for spintronic applications because of its supposedly long spin relaxation times\cite{Huertas2006, Han2014}. This is due to the low atomic number of carbon and the planar structure, which result in weak spin-orbit coupling compared to other conductors \cite{Gmitra2009, Kochan2017}. Calculations predicted graphene to have spin-lifetimes exceeding hundreds of nanoseconds \cite{Huertas2006, Ertler2009, Dora2010, Jeong2011, Dugaev2011, Pesin2012}. However, experiments up to now could only produce spin-lifetimes of a few nanoseconds \cite{Droegeler2014, Droegeler2016, Guimaraes2014, Singh2016}. Because of this discrepancy between theory and experiment, there has been an ongoing discussion of what is limiting spin-lifetimes in graphene. To increase the spin-lifetime in graphene, it is necessary to understand the limiting mechanisms to be able to design effective countermeasures. Several sources for additional spin relaxation in graphene have been proposed \cite{Han2014}: impurities (adatoms)\cite{Castro2009}, the substrate\cite{Ertler2009}, polymer residues\cite{Volmer2015, Avsar2016, Gurram2016}, ripples\cite{Huertas2009}, resonant magnetic scattering at magnetic impurities\cite{Kochan2014, Kochan2015} and contact induced spin relaxation\cite{Maassen2012, Volmer2013, Volmer2014}. To determine which is the dominant effect, experiments focused on finding a correlation between the momentum scattering time $\tau_p$ of electrons and their spin relaxation time $\tau_s$ \cite{Jozsa2009, Zomer2012, Swartz2013}. This would allow to differentiate between Elliott-Yafet type scattering, where $\tau_s \sim \tau_p$, and Dyakonov-Perel type scattering, where $\tau_s \sim 1/\tau_p$. This approach has so far produced no conclusive results \cite{Han2014}.

Another signature of spin relaxation mechanisms is the anisotropy or isotropy of the spin relaxation time. The out-of-plane spin relaxation time, which we will call $\tau_z$, can be different from the in-plane spin relaxation time, which we will call $\tau_{xy}$. For convenience, we introduce $\zeta:=\frac{\tau_z}{\tau_{xy}}$. 
For graphene on $\text{SiO}_2$ and prevailing Rashba type spin orbit fields $\zeta=0.5$ is expected \cite{Han2014}. In contrast, for resonant scattering at magnetic impurities, $\zeta=1$\cite{Kochan2014}. 
If contact induced spin relaxation is dominant, the spin relaxation measured in a Hanle experiment will also be isotropic.

\begin{figure}
\includegraphics[scale=0.22]{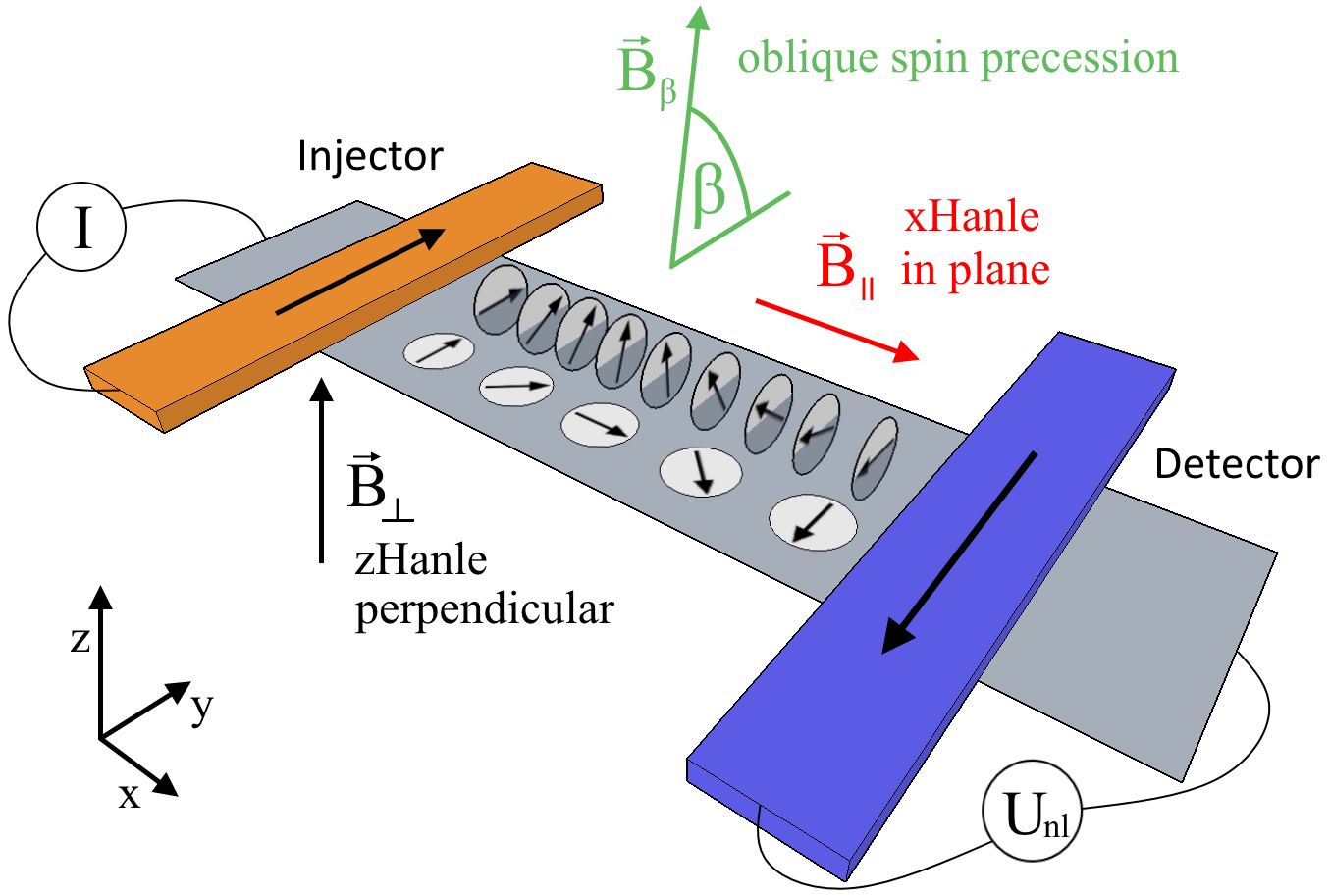}
\caption{\label{fig:schematic} Sample schematic illustrating the different orientations of the magnetic fields. {The non-local detection scheme, where the charge current path is outside the detector circuit removes spurious effects. The conventional zHanle experiment (black) rotates the spins only in the $x$-$y$-plane. The oblique spin precession experiment (green) was introduced in Ref.~\onlinecite{Raes2016}. In the xHanle experiment (red) the spins also experience the relaxation time $\tau_z$.}}
\end{figure}
\begin{figure}
\includegraphics[scale=0.8]{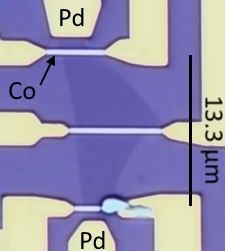}
\caption{\label{fig:microscope} Optical micrograph of the graphene flake with contacts. {Cobalt electrodes (light gray) serve as ferromagnetic injectors and detectors. Pd electrodes (yellow) provide the spin-independent reference probes and also contact the Co electrodes for AMR measurements.}}
\end{figure}

So far, experiments using two different methods to measure the anisotropy of the spin lifetime {in {pristine} graphene} have been published. We will discuss these methods and present experimental data of a third method. In the first experiment\cite{Tombros2008}, Hanle measurements were performed in a non-local geometry, with the magnetic field in $z$-direction (perpendicular to the graphene plane). A magnetic field in $z$ of up to two Tesla was applied to rotate the electrode magnetization from in plane to out of plane. The difference of the non-local signal at $B=0$ (magnetization in plane) and $B=2$ T (magnetization out of plane) was ascribed to the anisotropy of the spin-lifetime.

Another experiment utilized 
oblique spin precession \cite{Raes2016,Raes2017}. Again, a Hanle measurement is performed in non-local geometry, with the magnetic field in $z$-direction. The magnetic field is then tilted towards the electrode axis ($y$ axis, see Fig. \ref{fig:schematic}), which generates an out-of-plane spin population. At a sufficiently large $B$ field, dephasing due to spin precession leaves only spins aligned along the magnetic field direction detectable. The non-local signal measured as a function of the tilt angle can then be fitted to extract the anisotropy of $\tau_s$.

A third method, presented here, we call xHanle. We perform a Hanle measurement with the magnetic field applied along the $x$ axis (see Fig. \ref{fig:schematic}), so that the spins precess not just in plane but also out of plane. This xHanle trace is then compared to the standard Hanle measurement that we call zHanle. For isotropic $\tau_s$, zHanle and xHanle give identical results. {The xHanle method was recently used to measure the strong anisotropy in graphene in contact with transition metal dichalcogenides \cite{Benitez2017,Ghiasi2017} and was also suggested as a possible alternative to the oblique spin precession method \cite{Raes2017}. }

\begin{figure}
\includegraphics[width=0.4\textwidth]{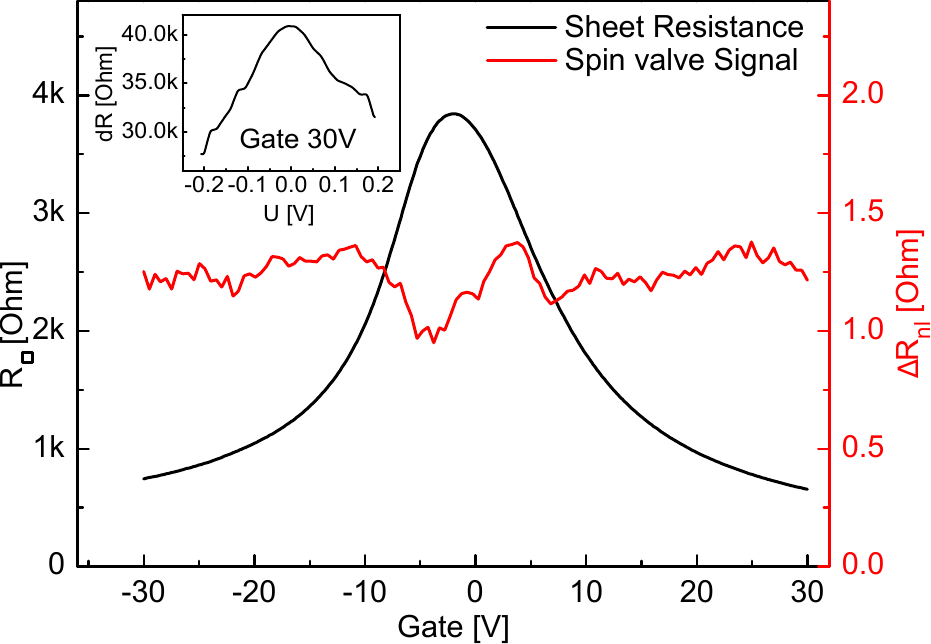}
\caption{\label{fig:gate} Gate dependence of the sheet resistance $R_\Box$ and the non-local spinvalve signal $\Delta R_{nl}$. The Dirac point is at -2 V. Inset shows the differential resistance of the injector tunnel contact as a function of current bias.}
\end{figure}
\begin{figure}
\includegraphics[width=0.4\textwidth]{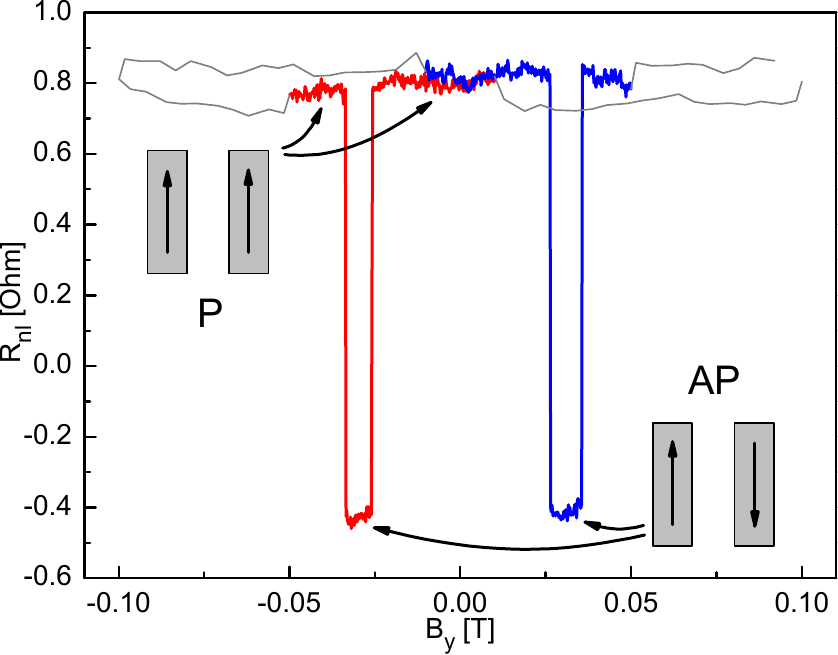}
\caption{\label{fig:spinvalve} Spin valve signal at $V_{bg} = 12$ V with illustrations to show the parallel (P) and antiparallel (AP) magnetic orientation of the electrodes. Distance of the injector and detector contacts was 13.3 $\mu$m with an injector current of 4 $\mu$A. {The grey trace shows the preparation of the electrodes that was done at a higher sweep rate, which induces an offset because of the DC measurement setup}.}
\end{figure}

\section{\label{sec:prep}Sample Preparation, Experimental Setup and Basic Characterization}

We use exfoliated single layer graphene on highly doped Si wafers (serving as a back gate) with 285\,nm $\text{SiO}_2$. The carrier mobility in graphene was about $\mu\approx 4000$\,$\text{cm}^2/\text{Vs}$. The electrodes are defined by electron beam lithography and electron beam evaporation of MgO/Co and Pd, respectively. The ferromagnetic Co electrodes are contacted by Pd leads on both ends to enable anisotropic magnetoresistance (AMR) measurements. The outermost contacts to the graphene sheet are also made of Pd, to have non-magnetic contacts which enable non-local spin valve measurements with only two switching contacts. To avoid the conductivity mismatch problem \cite{Schmidt2000}, we use a 1.4 nm thick MgO film underneath the magnetic Co contacts, having area resistances between 13\,k$\Omega\mu\text{m}^2$ and 46\,k$\Omega\mu\text{m}^2$. Fig. \ref{fig:microscope} shows a microscope image of the finished sample with five contacts (two Pd end contacts, three Co electrodes).

The measurements were performed in a cryostat with 3D vector magnet that consists of three superconducting magnetic coils, one large coil for the $z$ axis and two identical smaller ones inside the $z$ coil for the $x$ and $y$ field. The xHanle experiment requires all three coils. The $z$ and $x$ magnet are used for zHanle and xHanle, while the $y$ magnet is needed to switch the magnetic orientation of the electrodes from parallel (P) to antiparallel (AP). zHanle and xHanle were measured using different coils, so we checked if calibration errors, sample misalignment and stray fields influence our data. We tested the calibration of the magnets with a Hall probe and also tested for magnetic hysteresis {in a separate run}. We found a hysteresis loop in the $x$ and $y$ magnet extending up to an applied field of 250\,mT that can cause remanent fields of up to 2\,mT. 
Therefore, unwanted stray fields of up to 2\,mT can be present in the $xy$ plane during the measurement and need to be taken into account when analyzing the data.
We also checked for a misalignment of the sample and the magnets and found the possible misalignment angle to be below $3^\circ$. For a more detailed analysis of the magnetic setup, see the Supplemental Material\footnote{See Supplemental Material at
[URL will be inserted by publisher] for fabrication details, magnet setup, electrode magnetization, fitting procedure and estimation of uncertainty}.

Fig. \ref{fig:gate} shows a backgate sweep of the graphene sheet resistance, with the Dirac point at $V_{bg}=-2$\,V, indicating low extrinsic doping. For this measurement, the outermost electrodes were used to bias the sample and the voltage drop was detected between the Co electrodes that are later used as injector and detector for the spin experiments. The inset in Fig. \ref{fig:gate} shows the differential resistance $dR = dV/dI$ of the injector contact. The non-ohmic behavior is an indication for high quality tunnel barriers.

Spin transport measurements were carried out in a non-local DC setup schematically shown in Fig. \ref{fig:schematic} at $T=100$\,K. Below this temperature, switching the electrodes into an antiparallel state produced inconsistent results, which can be attributed to incomplete switching of the electrodes. 
Fig. \ref{fig:spinvalve} shows a spin valve measurement at 100\,K with properly switching electrodes and a spin valve signal of about $\Delta R_{nl}=1.2\,\Omega$.

The red graph in Fig. \ref{fig:gate} displays the gate dependence of the spin valve signal at an injector current of 4\,$\mu$A, used for all spin experiments here. The graph shows that the spin signal depends only weakly on $V_{bg}$. 
At negative injector bias there exists a regime where the back gate can be used to change the spin polarization of the injector current. This will be addressed elsewhere \footnote{{S. Ringer, M. Rosenauer, T. V\"olkl, M. Kadur, F. Hopperdietzel, J. Fabian, D. Weiss, J. Eroms, unpublished.}}. 
The experiments discussed in this paper were done at an injector bias where no change in the polarization of the injected spins occurs.

\begin{figure}
\includegraphics[width=0.4\textwidth]{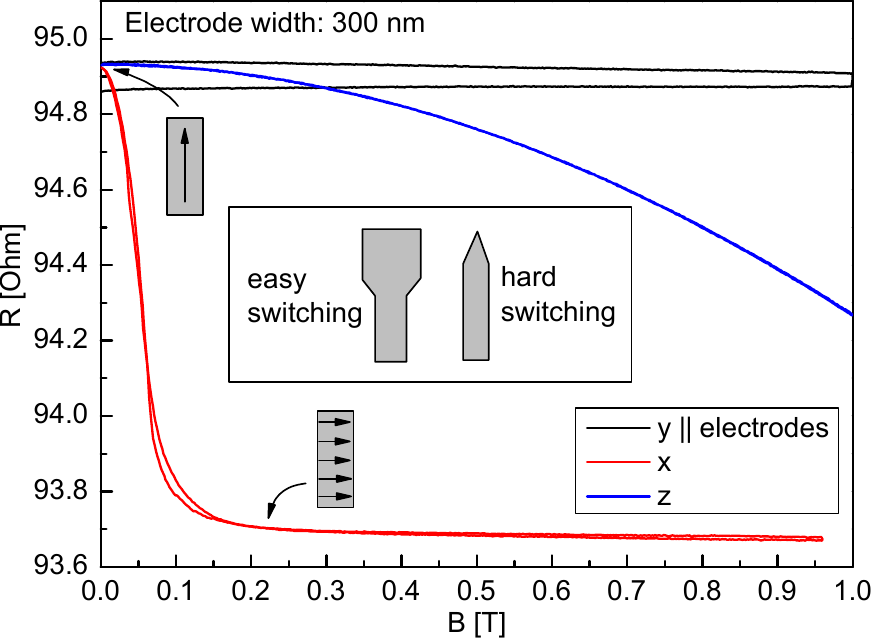}
\caption{\label{fig:AMR} AMR data of Co electrodes with the external field applied in $x$, $y$ and $z$-direction. Illustrations show the orientation of the magnetization in the electrodes. Inset: shape of the electrode tips for achieving different coercive fields while using the same width elsewhere.}
\end{figure}
For the xHanle measurement, it is essential to prevent the Co electrodes from rotating their magnetization. This places a limit on the maximum magnetic field that can be applied in $x$ direction. Narrow electrodes increase the magnetic shape anisotropy that keeps the magnetization aligned to the long axis. Additionally, a large distance between the contacts narrows the Hanle curve, reducing the required magnetic field.

It is common practice to use electrodes of different width to achieve different coercive fields needed to enable antiparallel switching of the electrodes. This is not practical for the xHanle experiment, as the electrodes should be as narrow as possible. Instead, we achieve different coercive fields by shaping the tips of the electrodes, as shown in Fig. \ref{fig:AMR}. A spatula-shaped tip reduces the coercive field, while preserving the magnetic stability with respect to perpendicular fields. Pointed tips, in contrast, increase the coercive field. 

AMR measurements were carried out to see at what field values the Co electrodes rotate. According to the AMR data displayed in Fig. \ref{fig:AMR},  at $B_x = 200$\,mT the electrodes are almost fully rotated into the $x$ direction. This rotation is independent of the tip shape, so the AMR data are the same for all electrodes. The peak width of the Hanle feature scales inversely with the travel time of the electrons. A long distance between  injector and detector contacts is therefore needed in order to narrow the Hanle feature to a field range well below 200\,mT. The xHanle measurements were done at magnetic fields only up to 25\,mT to avoid rotation of the electrodes. Corresponding measurements are presented in the Supplemental Material\cite{Note1}. For an injector-detector distance of 13.3\,$\mu$m most of the Hanle feature was in that field range.

To analyze the influence of an external field in arbitrary direction, including stray field, misalignment, and the anisotropy of spin relaxation, we employ the diffusion equation for the spin density $\vec s$:\cite{Fabian2007}
\begin{equation}
\frac{\partial \vec{s}}{\partial t} = \vec{s}\times\vec{\omega} + D\frac{\partial^2 \vec{s}}{\partial x^2} - {{\tau_s^{-1}}}{\vec{s}}
\end{equation}
with
\begin{equation}
{\tau_s^{-1}}= \begin{pmatrix} \tau_{xy}^{-1} & 0 & 0 \\ 0 & \tau_{xy}^{-1} & 0 \\ 0 & 0 & \tau_z^{-1} \end{pmatrix}
\end{equation}
the anisotropic spin relaxation rate, $D$ the spin diffusion constant and $\vec \omega$ the Larmor precession frequency vector, which is parallel to the magnetic field vector. While for Hanle experiments in isotropic media an analytical solution exists that is commonly used to fit the data \cite{Fabian2007}, we resort to a numerical finite element solution using the commercial software package COMSOL to account for anisotropic spin-lifetimes. {A detailed description of the COMSOL model is provided in the Supplemental Material\cite{Note1}. The simulated traces are then compared with experimental data of zHanle and xHanle. The parameters of the simulation are varied until the best possible match is obtained.}

\begin{figure}
\includegraphics[width=0.4\textwidth]{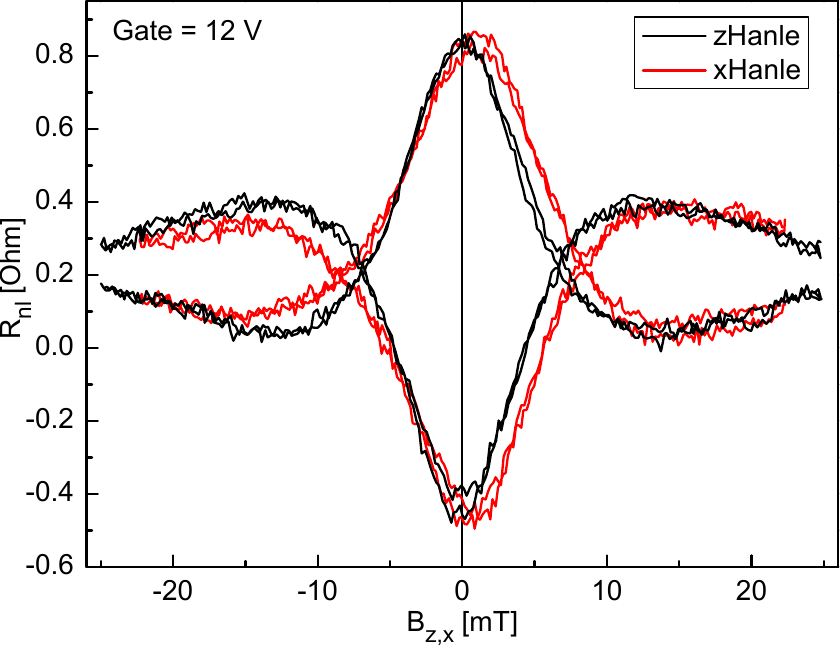}
\caption{\label{fig:xzHanleRaw} Raw data of zHanle (black) and xHanle (red). {Traces for both parallel and antiparallel magnetization were taken in both sweep directions.}}
\end{figure}

All Hanle measurements from which the spin-lifetimes were extracted were performed using the outer Co electrodes, which have a distance of 13.3\,$\mu$m, and an injector current of 4\,$\mu$A. At that distance, non-local spinvalve signals  $\Delta R_{nl}$ of 1.0 - 1.4\,$\Omega$ (depending on backgate voltage, see Fig. \ref{fig:gate}) could be achieved.

Fig. \ref{fig:xzHanleRaw} shows the raw data of zHanle (black) and xHanle (red) at $V_{bg} = 12$\,V. As can be seen, there is a distinctive difference between the traces, which will be discussed in more detail in subsection \textbf{III B}. The remanent magnetization in the magnet setup leads to a slight shift of the xHanle curve and has to be included for fitting of both zHanle and xHanle. 
For example, the small stray field in $x$-direction of about $B_x=0.8$\,mT that causes the xHanle center peaks to be shifted slightly off zero field also causes a small reduction to the height of the center peak of the zHanle trace.

\begin{figure}
\includegraphics[width=0.4\textwidth]{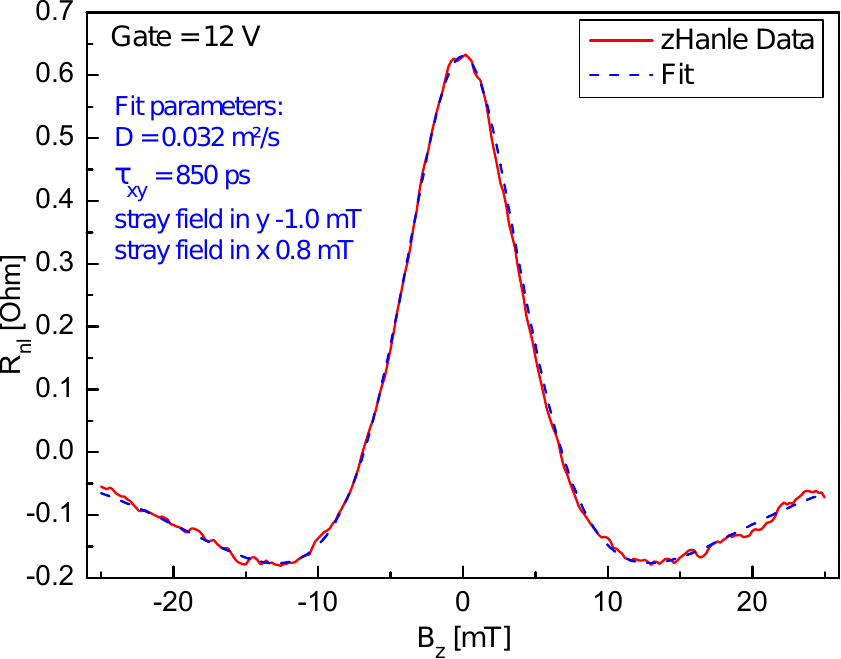}
\caption{\label{fig:zHanleFit} Smoothed data of zHanle (red trace) with fit for a stray field in $B_y=-1$ mT and $B_x=0.8$ mT (blue dashed line). {Smoothed data were obtained by averaging up and down sweep and subtracting P from AP sweep.}}
\end{figure}

We will first analyze the zHanle data, which is the standard characterization method in spin transport experiments. Fig. \ref{fig:zHanleFit} shows the smoothed zHanle data (obtained by averaging the up and down sweep, subtracting the AP signal from the P signal and dividing by two). Due to the presence of unknown stray fields up to about 2 mT, we also included a $B_y$ field during the fitting procedure. The best match to the data was obtained with a stray field of $B_x=0.8$\,mT  and $B_y=-1$ mT. 

\begin{figure}
\includegraphics[width=0.4\textwidth]{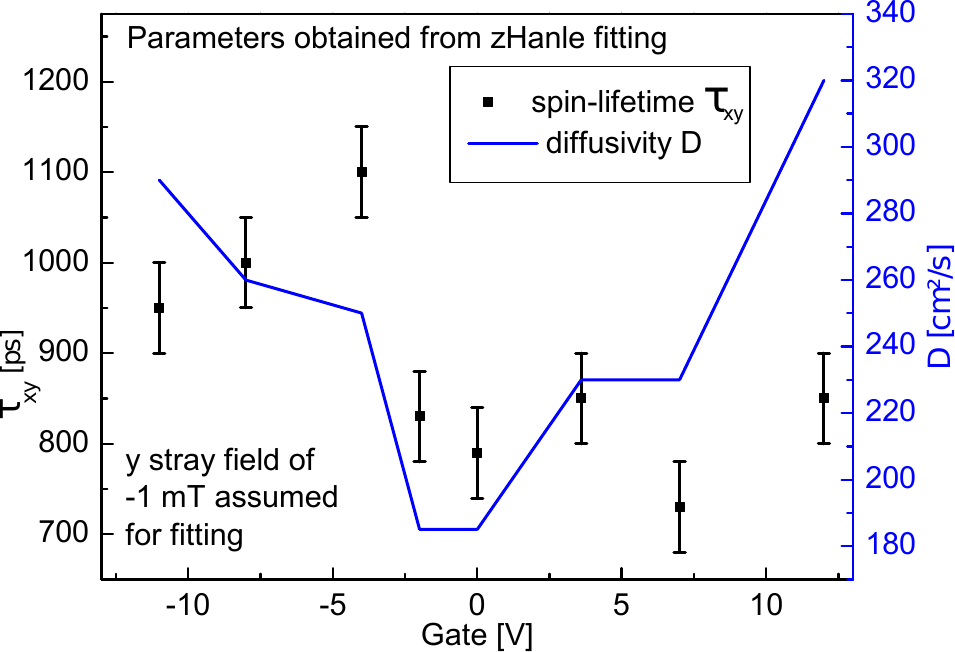}
\caption{\label{fig:fitparameter} In plane spin-lifetime $\tau_{xy}$ and diffusivity $D$ vs. gate voltage, extracted from fitting zHanle data with -1\,mT stray field in $y$.}
\end{figure}
We fitted the zHanle data for several gate voltages and extracted the parameters for spin diffusivity and spin-lifetime. 
Fig. \ref{fig:fitparameter} shows the fitted in plane spin-lifetime $\tau_{xy}$ and diffusivity $D$ plotted against the backgate voltage. The spin-lifetimes range from 730\,ps to 1100\,ps and show no correlation with the gate voltage. 
The spin diffusivity was a free parameter for the zHanle fit, giving 185\,$\text{cm}^2/\text{s}$ as the lowest value at the Dirac point and 320\,$\text{cm}^2/\text{s}$ as the highest value at $V_{bg} = 12$\,V. We also extracted $D_e$, the electron diffusivity, from the charge transport measurements shown in Fig. \ref{fig:gate}. Theses values are lower than the spin diffusivity ($D_e$ = 235\,$\text{cm}^2/\text{s}$ at $V_{bg} = 12$\,V).
Using the electron diffusivity as a fixed parameter for zHanle fitting produced significantly worse fits, so this was disregarded. {Note that due to the interdependence of the fitting parameters $D$ and $\tau_{xy}$ the error bars and spread are rather large.}

\begin{figure}
\includegraphics[width=0.4\textwidth]{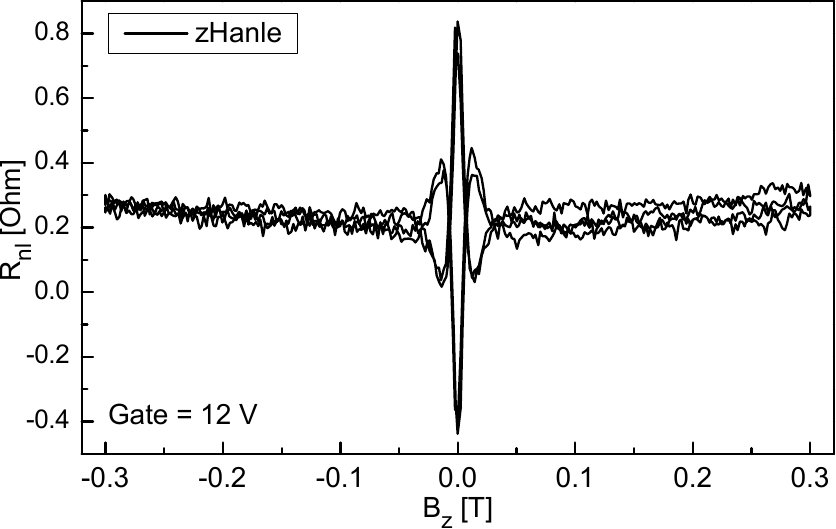}
\caption{\label{fig:zHanle300mT} zHanle measured up to 300\,mT (raw data) to see the background at higher fields {in parallel and antiparallel configuration}.}
\end{figure}
\begin{figure}
\includegraphics[width=0.4\textwidth]{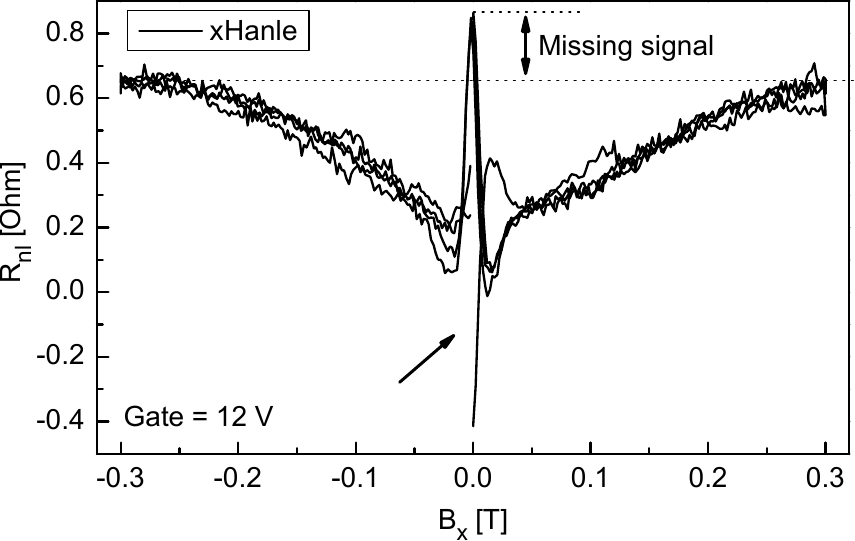}
\caption{\label{fig:xHanle300mT} xHanle measured up to 300\,mT (raw data) to see the rotation of the electrode magnetization. The upper double arrow indicates the difference between actual and expected signal at fields above 200\,mT. The lower arrow indicates the missing data of the AP downsweep, because the high $\text{B}_x$ field flipped the electrodes back to P.}
\end{figure}

\section{\label{sec:expts}Experiments on Anisotropic Spin Relaxation}
\subsection{\label{sec:rot}Rotating the Electrode Magnetization}

To check our setup, we performed zHanle and xHanle up to 300\,mT. The slight symmetric increase of the background in the zHanle data shown in Fig. \ref{fig:zHanle300mT} can be attributed to the Co electrodes slowly rotating into the external field towards the $z$ direction. We cannot fully rotate the electrodes towards $z$ as our magnet is limited to 1\,T. 
According to the AMR data in Fig. \ref{fig:AMR} however, 300\,mT is enough to rotate the electrodes completely towards the $x$ direction.
In this case, the injected spins should remain in plane and propagate without precession. Since $B_z$ remains zero, no orbital magnetoresistance effects that could possibly influence the detected signal should be expected. Therefore, we expect that for complete rotation of the electrodes towards $x$, the xHanle signal fully recovers the zero field {parallel} state value.\\
The 300\,mT xHanle is shown in Fig. \ref{fig:xHanle300mT} where the non-local signal at 300\,mT is noticeably smaller than the zero field value. Since at high $B_x$ the spin orientation remains in the graphene plane all the time, no anisotropy of the spin-lifetime is expected, so the signal loss must have a different origin. The most likely explanation for the signal loss is an imperfect magnetic alignment at the MgO-Co interfaces. Contrary to what the AMR data in Fig. \ref{fig:AMR} suggest, the interface magnetization is probably not yet fully aligned to the external field at 300\,mT. 
Spin injection and detection are sensitive to the interfaces of the electrodes, while AMR probes the bulk magnetization. It is known that a MgO-Co interface induces a strong magnetic coupling on the neighboring Co layers \cite{Yang2011}. This coupling seems to  make the interface magnetization more resistant to rotation than the bulk. More measurements that support this finding are discussed in the Supplemental Material\cite{Note1}.\\
We conclude that AMR data alone are not sufficient to characterize the electrodes for spin experiments. {A magnetic coupling at the electrode interface also exists in other common material combinations like AlOx-Co \cite{Monso2002,Rodmacq2003}. This needs to be considered when using the electrode rotation technique to determine the spin-lifetime anisotropy. Hence, in addition to a $B_z$ dependent background due to, \textit{e.g.}, the magnetoresistance of graphene \cite{Guimaraes2014,Raes2016}, the interface magnetization is another possible source of error and has to be taken into account. }

\subsection{\label{sec:xHanle}xHanle}
The xHanle experiment 
requires no rotation of the electrodes and no high magnetic fields. To extract the spin-lifetime anisotropy, we first determine $\tau_{xy}$ from the zHanle data as detailed at the end of section \textbf{II}.
For xHanle, the magnetic field is aligned along the $x$ axis (see Fig. \ref{fig:schematic}), and the spins precess in the $y$-$z$ plane. Therefore, the xHanle trace is sensitive not only to $\tau_{xy}$, but also to  $\tau_z$. For isotropic spin-lifetimes, xHanle and zHanle should give identical results.\\
We now discuss in more detail the raw data of zHanle (black) and xHanle (red) at $V_{bg}$ = 12\,V shown in Fig. \ref{fig:xzHanleRaw}. Both traces are not identical, which could be caused by anisotropic spin lifetimes. 
Also, we notice a clear asymmetry with respect to $B=0$ in the xHanle trace, not present in the zHanle data. This could be due to sample misalignment in combination with stray fields. We simulated Hanle curves for this situation (see Supplemental Material for details\cite{Note1}), and found that a sample rotation error of more than $12^\circ$ would be required to produce an asymmetry of the observed magnitude. As this is far more than our estimated error of $3^\circ$, and the resulting trace does not match the shape of our data, we disregard this scenario.

\begin{figure}
\includegraphics[width=0.4\textwidth]{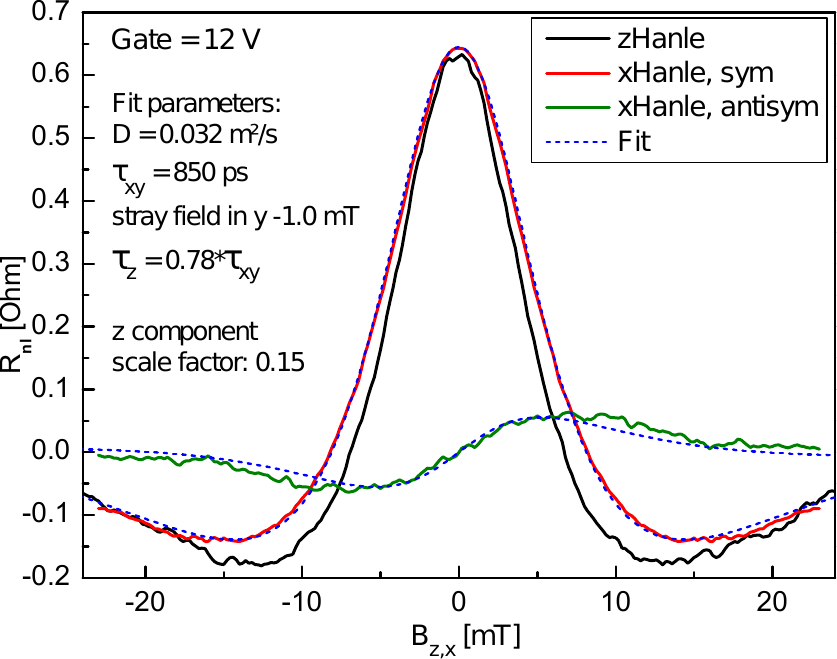}
\caption{\label{fig:FitxHanleSym} {Smoothed} zHanle data (black) and symmetrized and antisymmetrized {smoothed} xHanle data (red {and green}), with fit traces for the xHanle (blue dashed lines).}
\end{figure}

The most likely cause for the asymmetry of the xHanle signal is then a magnetization misalignment of the electrodes. Since the zHanle curve is \textit{not} asymmetric, we conclude that the magnetization of the detector electrode contains a small $z$ component in addition to the $y$ component expected from shape anisotropy. Considering that the Co electrodes have a film thickness of 20\,nm and are deposited on a near perfectly flat Si wafer, this tilted magnetization must be a local effect at the MgO-Co interface. It is know that a Co-MgO(100) interface induces a large perpendicular magnetic anisotropy in the neighboring Co layers \cite{Yang2011}. This anisotropy is heavily dependent on crystallinity and oxidization state. Both parameters are unknown.\\
To extract the data caused by the $y$ and $z$ components of the magnetization, we symmetrize and antisymmetrize the {smoothed} curves with respect to $B=0$. The result is shown in Fig. \ref{fig:FitxHanleSym}. 
We get a large symmetric part that is the projection on the $y$ component of the electrode magnetization and a smaller asymmetric part for the projection on the $z$ component of the magnetization. For isotropic spin relaxation, the symmetric part would be identical to the zHanle as the zHanle is also projected on the $y$ component of the magnetization.
The remaining difference between zHanle and the symmetrized xHanle is now due to the anisotropy in spin relaxation. The symmetrized xHanle can be fitted very well with an anisotropy of $\zeta= 0.78$. 
The antisymmetrized xHanle is fitted with the same parameters using only the scaling factor as a free variable. The scaling factor can then be used to estimate the tilt angle between injection and detection magnetization, which is $\sim9^\circ$.

Summarizing this subsection, we note that the xHanle experiment not only yielded the anisotropy $\zeta = 0.78$, but allowed to us to detect a small degree of $z$-magnetization in the electrodes.

\begin{figure}
\includegraphics[width=0.4\textwidth]{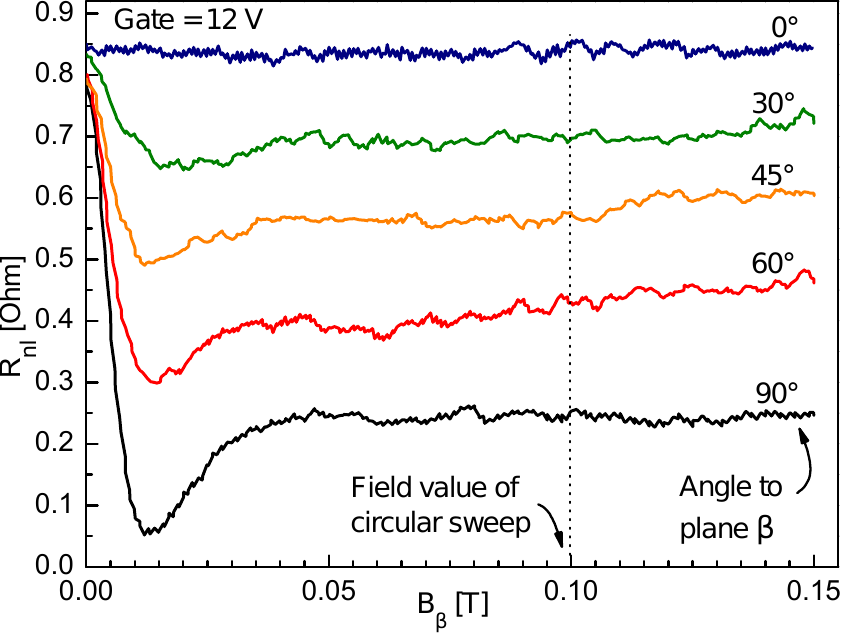}
\caption{\label{fig:WinkHan}Oblique spin precession traces at various inclination angles $\beta$ of the magnetic field. The data at $\beta = 90^\circ$ correspond to the zHanle experiment.}
\end{figure}
\begin{figure}
\includegraphics[width=0.4\textwidth]{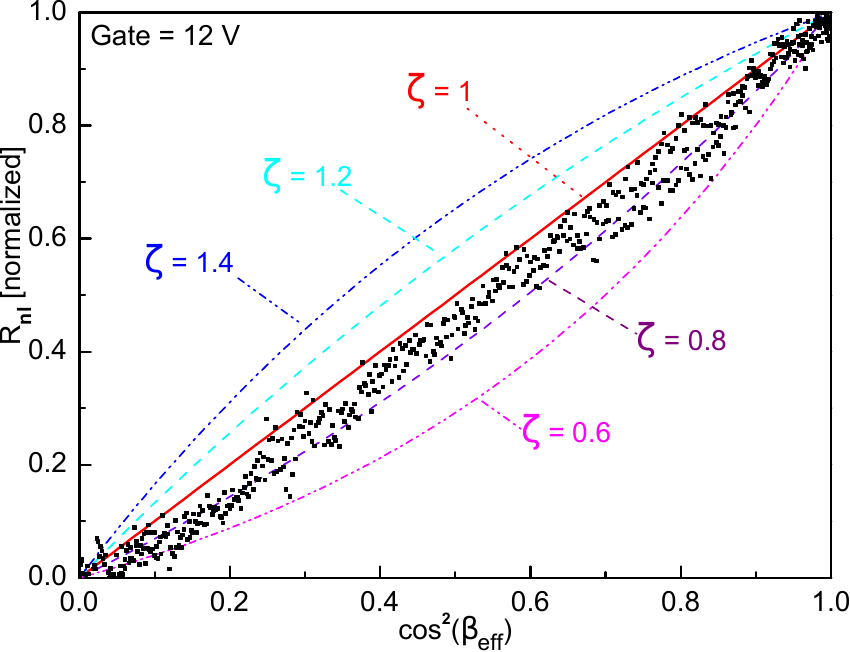}
\caption{\label{fig:Circ}Sweep of the field angle $\beta$ in the $z$-$y$ plane at a constant field of 100\,mT, plotted vs. $\cos^2\beta_\textrm{eff}$ to see the deviation from isotropic spin-lifetimes that is linear in this plot. Colored lines show the simulated traces for various degrees of anisotropy.}
\end{figure}

\subsection{\label{sec:oblique}Oblique Spin Precession}

Finally, we performed the oblique spin precession experiment of B. Raes \textit{et al.}\cite{Raes2016} on our sample. 
As outlined in the introduction, in this experiment an external field is applied at an angle $\beta$ to the $y$-axis (see also Fig. \ref{fig:schematic}).
When the field strength is varied at fixed $\beta$, we obtain a set of Hanle curves, shown in Fig. \ref{fig:WinkHan}. At large enough field strength, the spin component perpendicular to the external field is fully dephased, leaving only the component parallel to the external field. The projection of the original spin direction onto the external field direction results in a $\cos \beta$ term in the signal. During diffusion to the detector electrode, the spins are subject also to the out of plane spin relaxation time, if $\beta \ne 0$. When entering the detector electrode, the spins are now projected onto the magnetization of the detector electrode, resulting in a further $\cos \beta$ term. For isotropic spin relaxation, the non-local signal at the detector is therefore expected to follow a $\cos^2{\beta}$-dependence, while $\zeta \ne 1$ will lead to a deviation from that behavior. In Fig. \ref{fig:Circ}, we plot the spin signal of a continuous sweep of the angle $\beta$ at a fixed total external field of 100\,mT. 
To account for differences between the actual electrode magnetization direction and the $y$-axis direction, the data in Fig. \ref{fig:Circ} are plotted vs. $\cos^2\beta_\textrm{eff}$, where $\beta_\textrm{eff}$ is the angle between external field and the electrode magnetization direction. The $\cos^2$-scaling allows identifying any deviation from isotropic spin-lifetimes easily. 
The colored lines show simulated traces for various degrees of anisotropy, applying Eq. (7) in Ref.~\onlinecite{Raes2016} (see Supplemental Material for the full expression\cite{Note1}). 
The magnetization direction in the injector and detector electrodes deviates from the $y$-direction, which would be expected from shape anisotropy, due to rotation of the electrode magnetization in the external field and the partial $z$-magnetization ascribed to the MgO/Co-interface, which we detected in the xHanle experiment. This leads to correction terms that enter into the $\cos^2 \beta_\textrm{eff}$ term. More details on fitting procedure and formula are discussed in the Supplemental Material\cite{Note1}.
As can be seen, the data fall roughly between the linear (isotropic) trace and the $\zeta= 0.8 $ trace. A fit of the data gives an anisotropy of $\zeta= 0.91$. 
Importantly, the $z$-magnetization component at the MgO/Co-interface could only be detected in the xHanle experiment, but is crucial for the correct determination of $\zeta$. Not accounting for the $z$ component would have given an incorrect anisotropy of $\zeta = 1.08$\cite{Note1}. The precision at which the tilt angle can be determined enters into the error bars of the oblique spin precession method.

\begin{figure}
\includegraphics[width=0.4\textwidth]{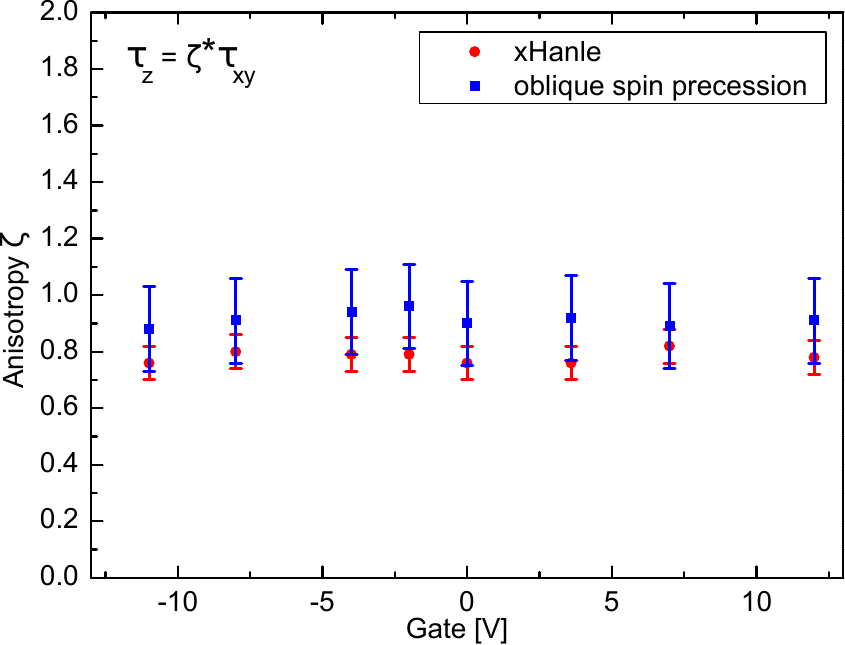}
\caption{\label{fig:Anisotropy}Extracted anisotropy ratio $\zeta$ from xHanle data (red) and oblique spin precession data (blue) as a function of gate voltage.}
\end{figure}

\section{\label{sec:disc}Discussion}
Fig. \ref{fig:Anisotropy} shows a comparison of the anisotropy extracted from the oblique spin precession experiment and the xHanle experiment. On average the xHanle experiment gives an anisotropy  $\zeta$ slightly below 0.8, while the oblique spin precession gives $\zeta$ a bit larger than 0.9. There is no correlation to the gate voltage.\\
Our values for $\zeta$ extracted from the oblique spin precession data are still in 
the range observed in previous experiments \cite{Raes2016}, with $\zeta$ between $\sim$0.9 and $\sim$1.03. The values extracted from our xHanle data are slightly below that range. \\
The xHanle experiment measures the P and AP configuration, while the oblique spin precession experiment  measures the P configuration only. First, this gives the xHanle experiment twice the amplitude over noise of the spin signal. Second, subtracting the AP from the P trace is a very reliable method to remove any background signal. The oblique spin precession experiment relies on normalization to remove any background.\\
The weakness of the oblique spin precession experiment is its sensitivity to the exact orientation of the electrode magnetization. On top of the permanent $z$ component caused by the MgO-Co interface, the dynamic tilting because of the external field needs to be accounted for in the fit formula\cite{Note1}. At our comparatively weak external field of 100\,mT, the dynamic tilting correction is responsible for a shift in $\zeta$ by roughly -0.07 and must be considered as a potential source of inaccuracy. Combined with the uncertainties related to determining the tilt angle of the permanent $z$ component, this leads to significant error bars for $\zeta$ extracted from fitting the oblique spin precession data.

It follows from those arguments that the xHanle experiment is generally more precise than the oblique spin precession experiment and also more robust to non-ideal conditions. As has been stated in section \ref{sec:oblique}, the permanent $z$ tilt in the detector electrode was only identified because of the xHanle experiment, but the knowledge of its existence was crucial for correct interpretation of the oblique spin precession data. We therefore conclude that the anisotropy in our sample is slightly below $\zeta$ = 0.8 as per the xHanle data, which is within the error bars of our oblique spin precession experiment, but in contrast to previous claims of an isotropic spin relaxation time\cite{Raes2016}. The disagreement between 
both results could be due to an overall higher spin relaxation rate in the experiment in Ref.~\onlinecite{Raes2016}.\\
For our sample, we assume that the spin relaxation stems from a mix of Rashba type spin-orbit fields that have a $\zeta$ of 0.5 and isotropic contributions like contact induced spin relaxation and resonant scattering by magnetic impurities. The individual relaxation rates are added to a total spin relaxation rate:
\begin{equation}
\frac{1}{\tau_{total}}=\frac{1}{\tau_{1}}+\frac{1}{\tau_{2}}+\ldots\quad .
\end{equation}
Assuming the anisotropic contributions are only of the Rashba type, we can use this formula to separate anisotropic and isotropic contributions. That gives us $\tau_{iso}=1.18$\,ns for the isotropic part and $\tau_{Rashba,xy}=3$\,ns and $\tau_{Rashba,z}=1.5$\,ns for the anisotropic Rashba part.\\
The spin lifetime of the isotropic part is consistent with the model of resonant scattering by magnetic impurities, such as adsorbed hydrogen\cite{Kochan2014, Kochan2015}, assuming a low concentration of scatterers. The sample was measured at a pressure of 
$\sim$10\,mbar, making a small concentration of hydrogen atoms or other species on the graphene surface plausible. This mechanism would exhibit a strong gate dependence when the energy approaches the resonance and no gate dependence at other energies. We do not see any significant gate dependence neither in the spin lifetime nor in the anisotropy {within the gate range of our experiment}. As the resonance can be outside of the energy range that we probed in our experiment ($E_F=116$\,meV at $V_{bg} = 12$\,V), this question could not be settled here.\\
Spin relaxation {induced by the ferromagnetic electrodes}, while still present, should be comparatively weak in our sample because of the high resistance area product of the contacts (cf. Fig. 5 in  Ref.~\onlinecite{Droegeler2016}). {The non-magnetic Pd electrodes have a lower contact resistance of about 600 $\Omega$. They are outside the spin transport path, but less than one spin-flip length away, so we assume a small contribution to isotropic spin relaxation. }\\

The local Rashba spin orbit fields caused by the few adatoms are not significant enough to matter. For the global Rashba spin orbit fields of the $\text{SiO}_2$ substrate, 
{initially, }C. Ertler \textit{et al.} calculated the spin relaxation time to be at least a few $\mu$s with a maximum at the CNP\cite{Ertler2009}. {Later, }  D. Van Tuan \textit{et al.} obtained a few hundred ps with a minimum at the CNP\cite{Tuan2016}. 
More recent DFT calculations suggest a Rashba spin-orbit coupling of $\lambda_R$ in the range of tens of $\mu$eV for graphene on crystalline SiO$_2$ \footnote{K. Zollner and J. Fabian, private communication}. Using the D'yakonov-Perel' mechanism, this works out to Rashba spin lifetime on the order of 1..10 ns, in agreement to our experimental data {and in line with calculations by Cummings and Roche for clean graphene \cite{Cummings2016}.}

\section{\label{sec:conc}Conclusion}
In conclusion, with the xHanle experiment we demonstrated an additional way to measure the anisotropy of the spin-lifetime in graphene that we believe to be so far the most accurate method. This tool can also be used to probe the spin relaxation in similar 2D materials that have recently started to attract interest like black phosphorus\cite{Avsar2017}. The data collected from the xHanle experiment pointed to a non-trivial magnetization of the contacts which is in line with the other experiments we performed. We attribute this non-trivial magnetization to a perpendicular magnetic anisotropy caused by the MgO-Co interface. Not accounting for this magnetization would have led to a false interpretation of the data from the oblique spin precession experiment. Compared to the oblique spin precession experiment, the xHanle is potentially more accurate, especially under non-ideal conditions. The weakness of this experiment is that it needs to operate at low magnetic field values to prevent a rotation of the electrodes. At these low field values, possible stray fields from the remanent magnetization of the magnets are relevant enough to influence the measurement.

The graphene sample in this study showed an anisotropy in the spin-lifetime of $\zeta$ slightly below $0.8$ that was clearly visible in xHanle but could not be identified with this precision in the oblique spin precession experiment. We conclude that the spin relaxation mechanism in our sample is a combination of isotropic and anisotropic parts. We attribute the isotropic part to resonant scattering at adatoms and also contact induced spin relaxation. The anisotropic part is due to Rashba spin orbit fields originating from the $\text{SiO}_2$ substrate.

\begin{acknowledgments}
Financial support by the Deutsche Forschungsgemeinschaft (DFG) within the programs SFB 689 (project A7) and  GRK 1570 is gratefully acknowledged, as well as financial support by the Elitenetzwerk Bayern. The authors would like to thank C. Back, G. Bayreuther, J. Fabian, D. Kochan and K. Zollner for fruitful discussions. The authors would also like to thank F. Volmer for his crucial advice on improving the quality of our MgO tunnel barriers and C. Strunk for the permission to use the UHV evaporator in his group.
\end{acknowledgments}


%



\section*{Supplementary material}

# Measuring anisotropic spin relaxation in graphene
# (Supplemental Material)

## I. CONTACTS AND TUNNEL BARRIER

The sample has electrodes with a width of 300 nm, which is a compromise of two conflicting requirements. More narrow electrodes are better for magnetic stability that is needed for xHanle, while wider electrodes are more reliable and allow for a thicker tunnel barrier at manageable contact resistances. A thick tunnel barrier is needed to reduce contact induced spin relaxation.

We used CSAR e-beam resist (Allresist AR-P 6200) for the lithography of the magnetic contacts and the tunnel barrier to reduce problems with polymer residues on graphene. Before the lithography, we annealed our sample at 200°C in vacuum for one hour as this improves the smoothness of the MgO film that will be deposited on the graphene. To increase the quality of the deposited MgO, we follow the advice of F. Volmer to keep the pressure of our deposition chamber below $1 \times 10^{-10}$ mbar and let our sample degas for at least a day in this pressure before material deposition.

## II. MAGNET SETUP

Our magnet setup consists of three superconducting magnet coils in a cryostat, one big coil for the $z$ axis and two identical smaller ones inside the $z$ coil for the $x$ and $y$ field. While the oblique spin precession experiment requires two coils, the xHanle experiment requires all three coils. The $y$ magnet is needed to switch the magnetic orientation of the electrodes from parallel to antiparallel. The $z$ and $x$ magnet are used for zHanle and xHanle, while $z$ and $y$ magnet together are used for the oblique magnetic field. zHanle, xHanle and the oblique spin precession were measured using different magnets, so we needed to check for calibration errors, sample misalignment and stray fields to properly analyse the data.

Our magnet setup has no permanently installed Hall sensor. After the measurements we checked the calibration of the magnets with a Hall probe and found that in the field range of 25 mT, the $x$ and $y$ magnets produced a field that was only 89% of what was expected. For larger fields the output was at 96%. The $z$ magnet was calibrated correctly. The data shown here use the correct $B$ field calibration, unless specifically stated otherwise. Because of this calibration error, the xHanle measurement range was actually 22 mT and not 25 mT.

We also tested for magnetization shifts at low field due to hysteresis, which is dependent on the magnetic field history of the magnets. We repeated the magnetic cycle that we used for the Hanle measurements and found offset fields of about 1-2 mT in the $x$ and $y$ magnet were possible. When the $x$ or $y$ magnet is set to zero, this offset remains present as stray fields and must be accounted for when processing the data.

The magnitude of the stray fields can be verified by fitting the zHanle data with various strength of stray fields. Fig. S1 shows two fit traces, one whith a stray field in $y$-direction of 1 mT and the other with 2.5 mT. As can be seen, the

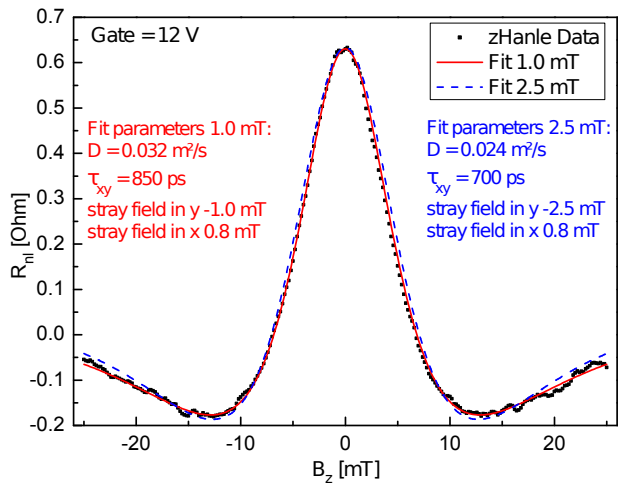

FIG. S1. Smoothed data of zHanle (black dots) with fit traces for a stray field in $y$ of -1 mT (red line) and -2.5 mT (blue dashed line).

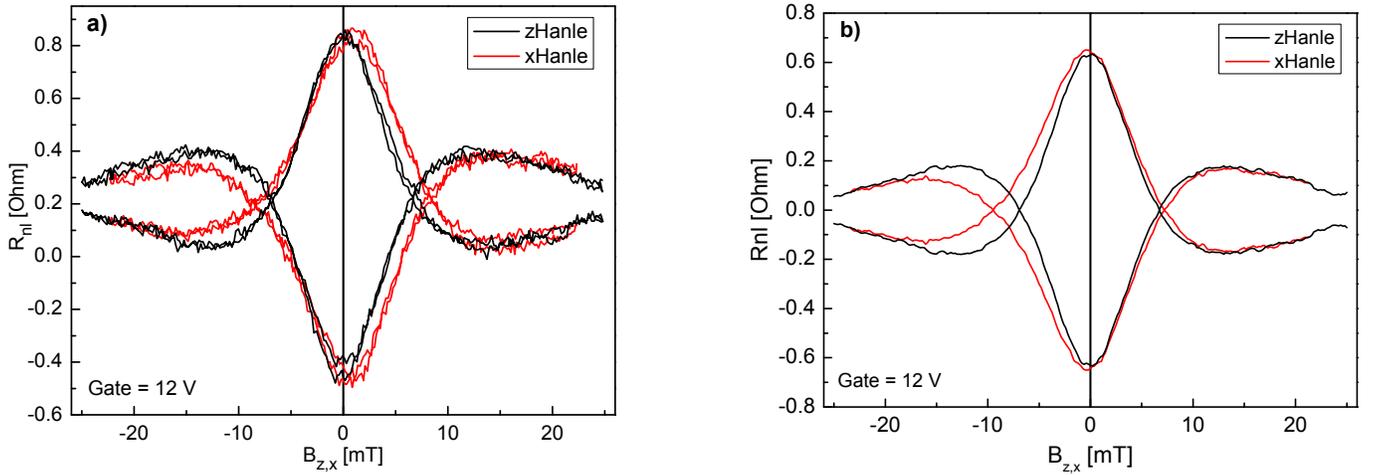

FIG. S2. a) Raw data and b) smoothed data of zHanle (black) and xHanle (red).

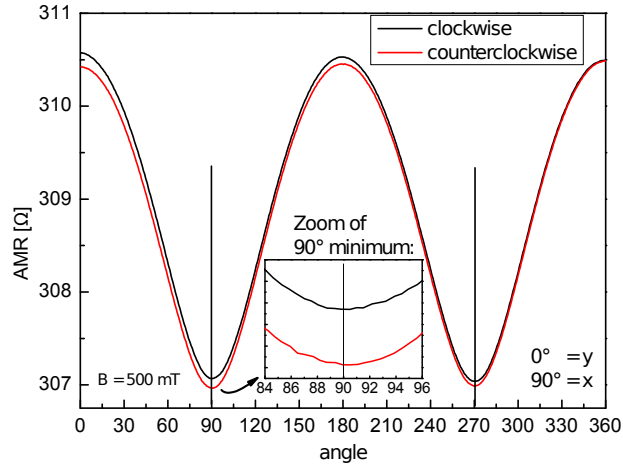

FIG. S3. AMR data of a circular sweep of the $B$ field in the $x$-$y$ plane at 500 mT, to test the alignment of the sample. Clockwise and counterclockwise sweeps were performed to account for possible magnetic hysteresis. A slight signal drift is visible because of temperature drift.

1 mT trace gives a perfect fit, while the 2.5 mT trace is noticeably worse in comparison. The shape of the zHanle data is therefore proof that the stray fields in $y$-direction are limited to 2 mT. For stray fields exceeding 1 mT, we note that the fit results for $D$ and $\tau_{xy}$ depend on the stray field strength, but this is not a concern for the assumed stray field of 1 mT.

Conclusions about remanent fields and hysteresis can also be drawn from looking at the raw data in Fig. S2a). Measurements were done as an up and down sweep, starting at 25 mT. There is no visible displacement between the up and down sweep, so magnetic hysteresis on this 25 mT loop can be neglected. The hysteresis loop that causes the $x$ and $y$ offset has a saturation field of 250 mT. The xHanle peak is shifted by 0.8 mT from zero field, which indicates a remanent magnetization of the $x$ magnet. This stray field in $x$ was also present during zHanle measurements and slightly reduced the peak amplitude. This is more obvious to see in Fig. S2b), where the up and down sweep is averaged, the background removed and the shift of the xHanle corrected. The $z$ magnet is more accurate and has a smaller remanent magnetization. The maximum observed shift of a zHanle peak was 0.4 mT (data not shown).

We also tested for sample misalignment by doing an AMR circle sweep in the $x$-$y$ plane, shown in Fig. S3. According to this data, the misalignment is about 1°. We swept in both directions to account for possible magnetic hysteresis. There was a slight drift in the AMR signal due to temperature drift. Including errors, we assume our sample was misaligned to the $B_x$ field direction by a maximum of 3°. The errors in this case is that the sample was taken out of the measurement setup after the xHanle data was collected and was remounted for the AMR circle sweep. The sample holder orientation is reproducible with an error of 2°.

## III. MAGNETIC ORIENTATION OF THE ELECTRODES

For a correct analysis of the xHanle data, it is important to know how exactly the electrodes rotate towards an external magnetic field in $x$ direction. According to the AMR data shown in Fig. S4, the electrodes do exhibit a slight rotation in the field range of $\sim 22\,\mathrm{mT}$ that is used for the xHanle measurement. To quantify this rotation, we look at the background signal of the xHanle, obtained by adding the P and AP trace. As can be seen in Fig. S5a), that background is essentially constant in that field range. A rotation of the electrodes would be indicated by an upwards inclination of the background signal with increasing $B$-field, as seen in Fig. S5b) at $\sim 40\,\mathrm{mT}$. There is no inclination visible in the background of Fig. S5a), indicating that in the field range of -22 mT to 22 mT the interface magnetization of the electrodes does not rotate. This is in disagreement with the AMR data, where the change in signal from 0 to 22 mT is about 7% of the total signal change, indicating a rotation of the bulk magnetization by $\sim 10°$.

As has been pointed out in the main article, the interface magnetization is what is important for spin injection and detection, while AMR probes the bulk magnetization. In our sample, the interface magnetization behaves slightly different from the bulk because of the magnetic coupling at the MgO-Co interface[1]. That we see no rotation of the electrodes in the xHanle background signal at -22 mT to 22 mT contrary to what we would have expected from the

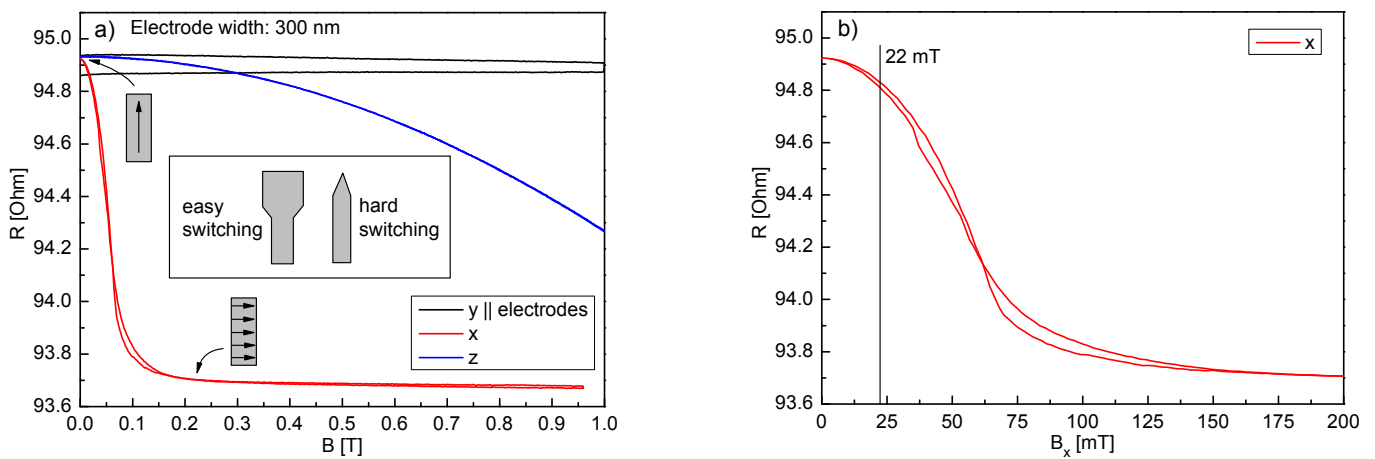

FIG. S4. a) AMR data of Co electrodes in $x$, $y$ and $z$. Illustrations show the orientation of the magnetization in the electrodes. Inset illustrates the shape of the electrode tips, to achieve different coercive fields while using the same body width. b) AMR data of the $x$ direction on a magnified scale.

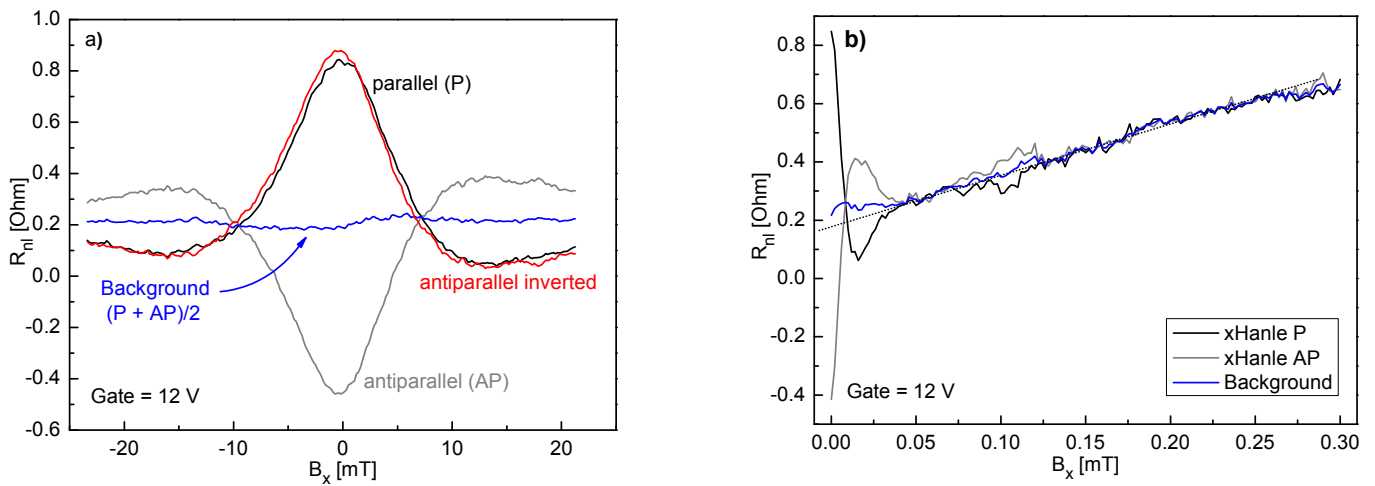

FIG. S5. xHanle data with extracted background ($\frac{P+AP}{2}$, blue) to check for possible rotation of electrodes. a) Field range of -22 mT to 22 mT with fine resolution. The red trace is the inverted AP signal for comparison with the P trace, to check for proper AP alignment, electrode stability, and a possible change in the stray fields caused by the switching. b) Field range of 0 to 300 mT with coarse resolution. The dashed line is a guide to the eye to underline the onset of electrode rotation at $\sim 40\,\mathrm{mT}$.

AMR data is more evidence to support this thesis.

Fig. S5a) also shows the flipped xHanle antiparallel trace, for a direct comparison with the parallel trace. First, this can be used to compare amplitudes of the P and AP signal to ensure that the electrodes have switched properly. As has been stated in the main text, incomplete switching can occur at temperatures below 100 K. Second, this was done to test if the remanent fields of the $y$ magnet may have changed because of the switching procedure. The traces are identical enough to assume unchanged stray fields.

Finally, to address the difference between the rotation of electrode magnetization extracted from AMR and xHanle, we simulated the expected xHanle curve, using anisotropic $\tau_s$, but including rotation of magnetization at both injector and detector, using the angle extracted from AMR. In Fig. S6 (left), we see that magnetization rotation leads to a visible upturn of the background curve (blue) and a difference between parallel (black) and flipped antiparallel (red) curve, in contrast to the experimental finding. We also checked if a slight electrode rotation would lead to a broadening of the xHanle curve even without anisotropy. As can be seen in Fig. S6 (right), despite the electrode rotation as extracted from AMR, the central part of the calculated isotropic Hanle curve is not broadened and coincides with the measured zHanle curve. Therefore, even when including some degree of electrode rotation, the xHanle experiment can still clearly distinguish between isotropic and anisotropic $\tau_s$.

## IV. HANLE FITTING

To prepare the data for fitting, we averaged the up and down sweep and subtracted the antiparallel from the parallel trace (then divide by two) to eliminate the background. The data was also shifted where necessary to have the center peak at zero field. A constraint for the fitting is that xHanle and zHanle need to be fitted with the same diffusivity $D$ and $\tau_{xy}$. We also assumed identical stray fields for both fits. The fitting procedure was then to first fit the zHanle to obtain these parameters and then use them in the xHanle fit.

As the result of B. Raes's oblique spin precession experiment[2] was that spin-lifetimes are isotropic, we first tried to fit the xHanle data by assuming isotropic spin-lifetimes while modeling the differences to the zHanle data with sample misalignment and stray fields in $y$. The obvious problem to fit the xHanle data is that the trace is asymmetric regarding the magnetic field direction. We found that a combination of stray fields in $y$ and a misaligned sample that is rotated in the $x$-$y$ plane can indeed produce an asymmetric xHanle trace. However, misalignment or a $y$ stray field alone do *not* produce an asymmetry, the combination of both is needed for that.

Fig. S7 shows a fit of the xHanle data with a parameter sweep of the misalignment angle, with a stray field in $y$ of 1 mT. Assuming isotropic spin-lifetimes, this is the closest we could get to fit the data. Higher stray fields in $y$ did not give a better fit. As can be seen, the data on the positive $B$ side can be fitted reasonably well, while the negative $B$ side gives no good fit at all even for unreasonably large misalignment angles. Large misalignment angles produce a trace with considerable asymmetry, but the shape does not fit. However, we measured the misalignment angle to be $3°$ at maximum as detailed in section II, and as the simulation shows, that angle is not enough to create an asymmetry of the magnitude seen in the xHanle data. Therefore, we conclude that the asymmetry in the xHanle

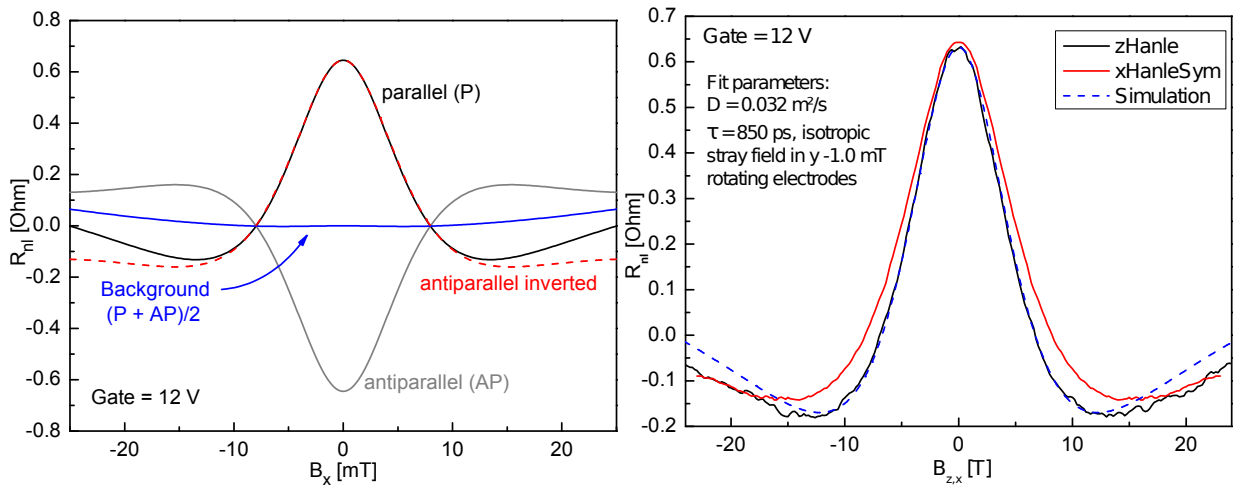

FIG. S6. Left: Simulated xHanle curve, using anisotropic $\tau_s$ and rotating electrode magnetization. Colors as in Fig. S5(a) Right: Simulated Hanle curve, including electrode rotation, but with isotropic $\tau_s$.

data can not be explained by stray fields and sample misalignment.

We also examined the effect of stray fields in $x$ for zHanle or in $z$ for xHanle. We found that these stray fields reduce the height of the center peak of the Hanle feature, but have nearly no effect on the shape of the secondary peaks.

Next, we did not require $\tau_z = \tau_{xy}$ but allowed $\tau_z$ as a free parameter for the xHanle fit. This produced a good fit of either the positive $B$ side data *or* the negative $B$ side data. A good fit of *both* sides would require a separate $\tau_z$ for each side. However, there is no known effect where the anisotropy changes when the magnetic field is reversed. The asymmetry of the xHanle data has a different origin.

The best explanation for the asymmetry is a $z$ component in the magnetization of one of the electrode interfaces. This $z$ component could be a tilted magnetization or actual magnetic domains with a $z$ orientation. It is not present in both electrodes, as that would result in a dip or peak in the spinvalve data. We cannot say which electrode has the $z$ component, but we assume for ease of understanding and writing that it is the detector electrode. As both electrodes have the same width and are deposited in the same lithography step, this raises the question of how can there be a difference. Ideally, they should be identical, but there is evidence that they are not. For example, the resistance area product of the two electrodes is quite different, one being at $13\,\mathrm{k\Omega\mu m^2}$ and the other at $46\,\mathrm{k\Omega\mu m^2}$. This difference is most likely caused by the inhomogeneity of the MgO tunnel barrier, where the current flows through local hot spots where the barrier is thinnest.

The MgO grows polycrystalline, and the Co on top of the MgO is then most likely polycrystalline as well. While the magnetization direction of the electrode interfaces can be expected to be the same on average, there are likely local fluctuations because the strength of the MgO-Co exchange coupling depends on the local crystal structure. Such a local fluctuation can get amplified disproportionately in the spin signal when it is at a tunneling hotspot. The $z$ component in our spin signal is then the result of statistical fluctuations.

How this $z$ component in the magnetization can cause an asymmetric Hanle signal is schematically shown in Fig. S8. For an xHanle experiment in isotropic media where the spins propagate in the $x$ direction with parallel electrodes in $y$ and a magnetic field in $x$, the spin signal at the detector electrode is proportional to:

$$s_y(x,t) \sim \int_0^\infty dt \frac{1}{\sqrt{4\pi Dt}} e^{-(x-\mu Et)^2/4Dt} e^{-t/\tau_s} \cos(\omega_0 t) \tag{1}$$

This formula is also commonly used to fit zHanle. The trigonometric function that defines the shape of the Hanle trace is a cosine, and accordingly we call this a cosine shaped Hanle as depicted in Fig. S8a).

Now we do the same xHanle in isotropic media, but the detector electrode is pointed in the $z$ direction. We detect the $z$ component of the spin that is described by the following equation:

$$s_z(x,t) \sim \int_0^\infty dt \frac{1}{\sqrt{4\pi Dt}} e^{-(x-\mu Et)^2/4Dt} e^{-t/\tau_s} \sin(\omega_0 t) \tag{2}$$

This formula describes Hanle where the injector and detector are not parallel but perpendicular to each other. The trigonometric function that defines the shape of this Hanle trace is a sine, and accordingly we call this a sine shaped Hanle as depicted in Fig. S8b).

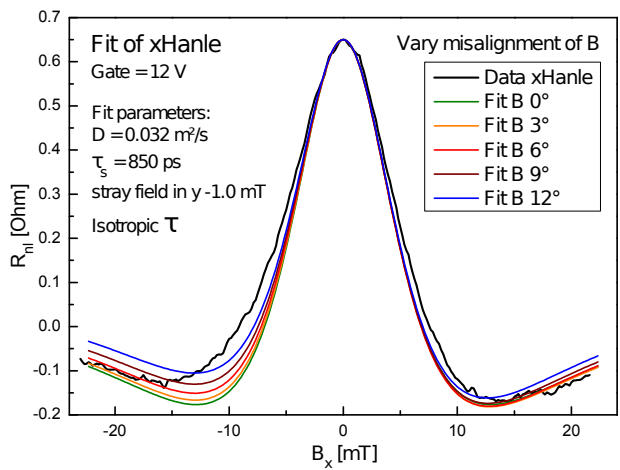

FIG. S7. xHanle data (black), with fit traces assuming isotropic spin-lifetimes. Parameter sweep of the sample misalignment angle.

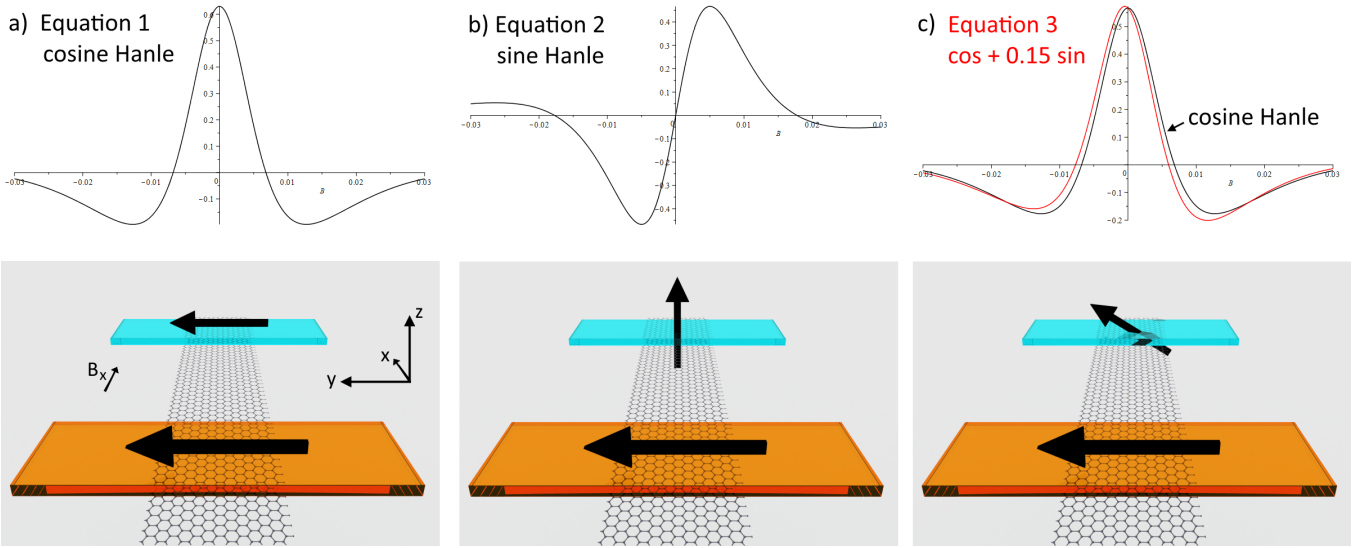

FIG. S8. a) cosine shaped Hanle, b) sine shaped Hanle, c) cosine shaped Hanle in black that is symmetric and in red a linear combination of cosine and sine Hanle that is asymmetric. Pictures below the graphs illustrate the corresponding orientation of the electrode magnetization, assuming an xHanle experiment. Rendered image courtesy of H. Maier.

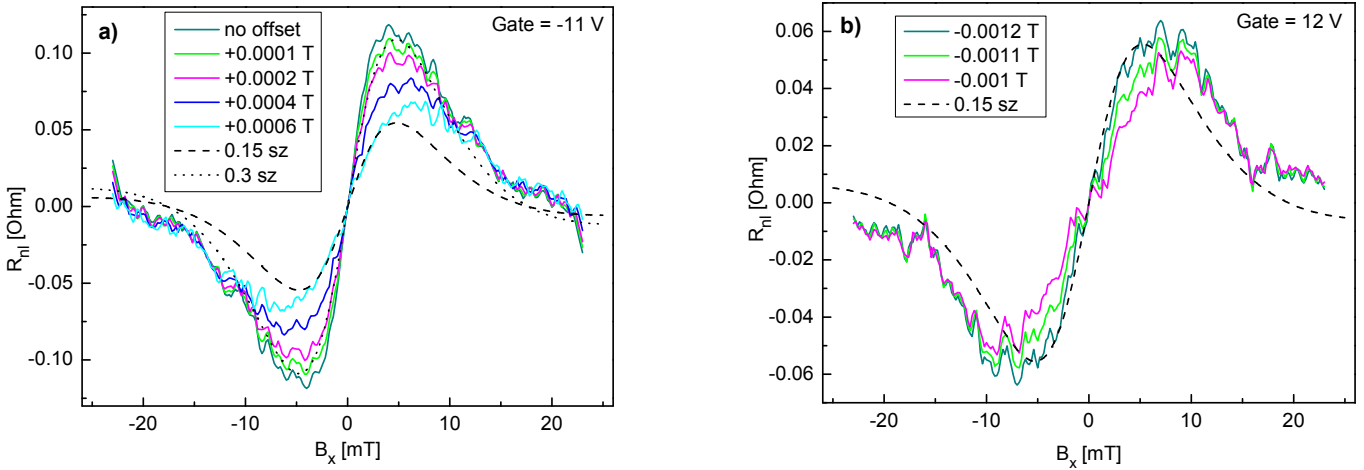

FIG. S9. Antisymmetrized part of the xHanle data, calculated from different offset fields to correct for remanent magnetization.

If injector and detector are neither parallel nor perpendicular but at an angle in between, then we have a linear combination of cosine and sine shaped Hanle. This mixed case is described by the following formula:

$$s_{y,z}(x,t) \sim \int_0^\infty dt \frac{1}{\sqrt{4\pi Dt}} e^{-(x-\mu Et)^2/4Dt} e^{-t/\tau_s}(\cos(\omega_0 t) + a \cdot \sin(\omega_0 t)) \qquad (3)$$

The tilt angle $\phi$ between injector and detector electrode is being represented by the scaling factor $a$, with $\phi = \arctan(a)$. A mixed Hanle trace with $a = 0.15$ and $\arctan(a) = 8.53°$ is shown in in Fig. S8c) in red, with a cosine shaped Hanle in black for comparison. As can be seen, the red trace is asymmetric.

Equation 1 is symmetric in $B$ while equation 2 is antisymmetric in $B$. The mixed Hanle of equation 3 that is a linear combination of the two can therefore be separated into these two parts by symmetrization and antisymmetrization. By symmetrizing the xHanle data we can extract the $y$ part of the xHanle and compare that to the zHanle to extract the anisotropy of the spinrelaxation. The zHanle data has also just a $y$ component. We first fit the zHanle to obtain the fit parameters, then the $y$ part of the xHanle is fitted with the same parameters and the only free variable is the anisotropy of the spin lifetime. To do that, we used a COMSOL simulation as detailed in the main text, as the formulas 1 - 3 do not account for an anisotropic spinrelaxation.

Antisymmetrizing the xHanle data gives the $z$ component of the spin signal that can be used to extract the scaling

factor $a$ and thus the tilt angle $\phi$ of the detector electrode magnetization. We use the same fitting parameters as for the $y$ component of the xHanle and fit with a sine shaped Hanle, allowing only $a$ as a free variable. It should be noted that knowing $a$ is *not* needed to extract the anisotropy out of the xHanle data. But it *is* needed for a correct interpretation of the oblique spin precession data.

Estimating the tilt angle from the antisymmetrized xHanle data is not very accurate because of our problems with remanent $B$ fields in the $x$ magnet as stated in section II. This is demonstrated in Fig. S9 that shows the antisymmetrized xHanle data for various small offsets to account for a possible remanent field. As can be seen in Fig. S9a), for no offset a scaling factor of 0.3 would give a good fit, while for an offset of 0.6 mT a scaling factor of 0.15 is the best possible fit. It must be concluded that for this kind of analysis, our magnet setup is not accurate enough for a reliable result.

As our measurements were reproducible, we can assume that the $z$ component of the interface magnetization and thus the scaling factor $a$ is a fixed parameter that does not change because of the gate or the external magnetic field. We considered the antisymmetrized data from all gate voltages under different offset corrections and tried to find offset fields were all data could be fitted with the same scaling factor $a$. We found that a scaling factor of 0.15 that corresponds to a tilt angle of 8.53° can fit all data reasonably well. We cannot say whether the $z$ component of the magnetization is a tilt of the whole magnetization or domains with a $z$ direction. Mathematically, they are identical. In equation 3, instead of $(\cos(\omega_0 t) + a \cdot \sin(\omega_0 t))$ which is the domain representation, one could also use $(\cos(\omega_0 t - \phi)/\cos(\phi))$, which would be the tilt angle representation. In case of domains, the percentage of z oriented domains is $\frac{a}{1+a}$ and the percentage of $y$ oriented domains is $1 - \frac{a}{1+a}$. For $a = 0.15$ this would mean that 13% of the domains that contribute to the spin detection are $z$ domains.

## V. OBLIQUE SPIN PRECESSION FITTING

To prepare the oblique spin precession data for fitting it needs to be normalized to a scale of 0 to 1. $\beta$ is the angle of the magnetic field to the $x$-$y$-plane. For $\beta = 90°$ the spin signal should be 0, so any remaining signal is background that needs to be subtracted from the data. The data is then divided by its peak value which is at $\beta = 0°$ to normalize it to 1. To fit the data, we used the formula provided by B. Raes[2]:

$$\frac{R_{nl}^{\beta}(B)}{R_{nl}(B=0)} = \sqrt{\left(\cos^2(\beta) + \frac{1}{\zeta}\sin^2(\beta)\right)^{-1}} \cdot e^{-\sqrt{\frac{L^2}{\tau_{xy}D}}\left(\sqrt{\cos^2(\beta)+\frac{1}{\zeta}\sin^2(\beta)}-1\right)} \cdot [\cos^2(\beta-\gamma)] \qquad (4)$$

Here, $\frac{R_{nl}^{\beta}(B)}{R_{nl}(B=0)}$ is the normalized non-local spin signal. The formula in Ref. 2 has an additional term for the magnetoresistance of the graphene that we omitted as we operate at a low enough magnetic field. $L$ is the distance of the injector and detector contact, $D$ is the diffusivity and $\zeta = \frac{\tau_z}{\tau_{xy}}$ is the anisotropy of the spin relaxation time. When the field component perpendicular to the plane increases, the electrodes will start to turn into the $z$ direction as shown in the AMR data in Fig. S4 (blue trace). In equation 4 this is accounted for by $\gamma$, which is the field dependent

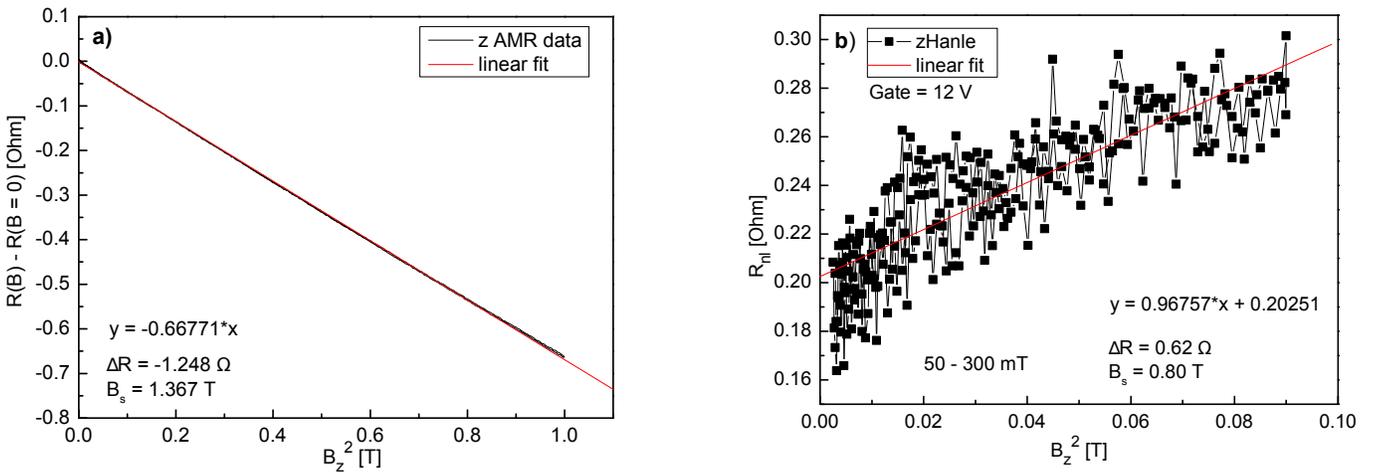

FIG. S10. a) $z$ direction AMR data plotted vs. the quadratic magnetic field. Linear fit in red. b) zHanle data of 50 - 300 mT plotted vs. the quadratic magnetic field. Linear fit in red.

angle of the symmetric turning of the electrodes in the $z$ direction because of the $z$ component of the external field. This $\gamma$ does *not* account for the $z$ component of the interface magnetization that exists in our electrodes independent of an external field. That $z$ component that we represent with a tilt angle $\phi$ is not yet accounted for in equation 4, but is added later in equation 7.

A formula to determine $\gamma$ is given in the supplemental material of Ref. 2:

$$\gamma = \arcsin[\frac{\sin(\beta)}{B_s/B + \cos(\beta)}] \quad (5)$$

Here, $B_s$ is the saturation field that is extracted from the AMR data using the following formula:

$$\rho(B) - \rho(B=0) = \frac{\Delta\rho}{B_z^2}B^2 \quad (6)$$

We use the AMR data where the field is applied along the $z$ axis, which is the blue trace in Fig. S4. Then $\rho(B)$ is the field dependent electrical resistance and $\Delta\rho$ is the maximum resistance difference of the AMR data. From the $y$ data (red) in Fig. S4 we see that $\Delta\rho = -1.248\,\Omega$. We plot $\rho(B) - \rho(B=0)$ vs. $B^2$ as shown in Fig. S10a) and do a linear fit, where the inclination of the linear fit is equivalent to $\frac{\Delta\rho}{B_z^2}$. So we have $-0.66771 = \frac{\Delta\rho}{B_z^2}$, which gives us $B_s = 1.367\,\mathrm{T}$.

However, AMR probes the bulk magnetization and the $B_s$ we extracted from the AMR data will give us a $\gamma$ that describes the rotation of the bulk. Since we know that because of the magnetic coupling at the MgO-Co interface, the interface magnetization behaves differently than the bulk, we can expect the interface to rotate differently than the bulk. We can see the rotation of the interface in our 300 mT zHanle data and can use that to extract a $B_s$ for the interface. In Fig. S10b) we plot the dephased zHanle signal vs. $B^2$ and do a linear fit. We used the zHanle data of field values above 50 mT to ensure a dephased signal. The $\Delta\rho$ is then equivalent to the amplitude of the center Hanle peak, which is $0.62\,\Omega$. The fit results in $B_s = 0.80\,\mathrm{T}$. The saturation field for the interface magnetization is lower than for the bulk, meaning the interface magnetization is easier to rotate into the $z$ direction. This is consistent with our assumption of the MgO-Co magnetic coupling at the interface, considering that the coupling favors an out of plane magnetization. To further process the data, we used $B_s = 0.80\,\mathrm{T}$.

We extended equation 4 to account for the $z$ component of the interface magnetization:

$$\frac{R_{nl}^\beta(B)}{R_{nl}(B=0)} = \sqrt{\left(\cos^2(\beta) + \frac{1}{\zeta}\sin^2(\beta)\right)^{-1}} \cdot e^{-\sqrt{\frac{L^2}{\tau_{xy}D}}\left(\sqrt{\cos^2(\beta)+\frac{1}{\zeta}\sin^2(\beta)}-1\right)} \cdot [\cos(\beta-\gamma)\cdot\cos(\beta-\gamma-\frac{\beta}{|\beta|}\phi)/\cos(\phi)] \quad (7)$$

Here, $\phi = \arctan(a)$ is the tilt angle that corresponds to a scaling factor of $a$ for the $z$ component. The term $\frac{\beta}{|\beta|}$ is there to change the sign of $\phi$, as the $z$ component switches direction when the external $z$ field switches direction. That this is the case can be seen in Fig. S11, where the oblique spin precession signal is plotted vs. $\beta$ for a range of $\beta = 90°$ to $\beta = -100°$. The measurement started at $\beta = 90°$ and at $\beta = 0°$ the $z$ component of the external field changes direction.

The continuous red fit trace in Fig. S11 uses equation 7, with $\phi = 8.53°$ and $\zeta$ as a free parameter. The dashed red trace also uses equation 7, but here $\phi$ does not change sign, meaning the $z$ component of the interface magnetization does *not* follow the external field. As can be seen, this does not fit the data. The conclusion is that the $z$ component of the electrode magnetization follows the external field and switches direction.

The blue trace is for comparison. It uses equation 4 that assumes no $z$ component in the electrode magnetization, again with $\zeta$ as a free parameter. The differences between the red and the blue fit trace are small, but it does seem that the red trace gives a better fit near $\beta = 0°$, supporting our tilted electrode thesis. It must be noted, however, that the way the data is normalized can change which trace seems to give the better fit.

The blue fit would give a higher $\zeta = 1.08$ compared to $\zeta = 0.91$ of the red fit. This demonstrates how big the influence of the $z$ component of the electrode magnetization is when fitting for the spin-lifetime anisotropy. As has been stated at the end of section IV, our method to estimate the magnitude of the $z$ component is not very accurate. This inaccuracy is passed on to $\zeta$ when fitting the oblique spin precession data. A larger scaling factor than 0.15 for the $z$ component would lower $\zeta$ while a smaller scaling factor would increase $\zeta$.

## VI. ESTIMATION OF UNCERTAINTY

The uncertainty of the parameter $\zeta$ calculated from the xHanle method is $\pm 0.06$ for an average $\zeta$ of 0.78 as displayed in Fig. 14 of the main text. There are two sources of uncertainty for the xHanle. The first is the uncertainty of the fit,

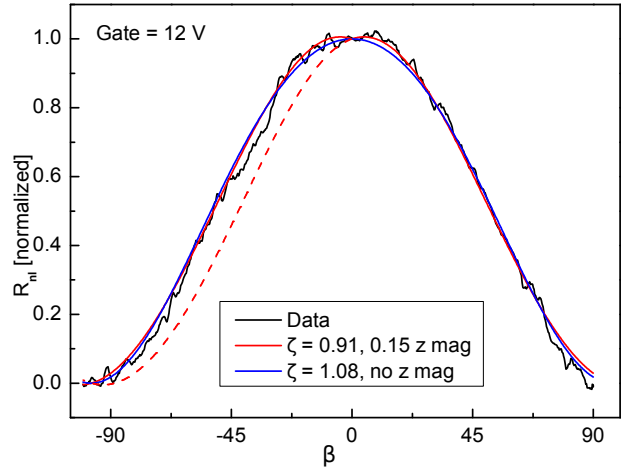

FIG. S11. Oblique spin precession data. The red fit assumes an interface magnetization $z$ component of 0.15 that switches direction as soon as $\beta$ changes sign. The dashed red line is the continued trace where the $z$ component does not switch direction. The blue fit assumes no $z$ component.

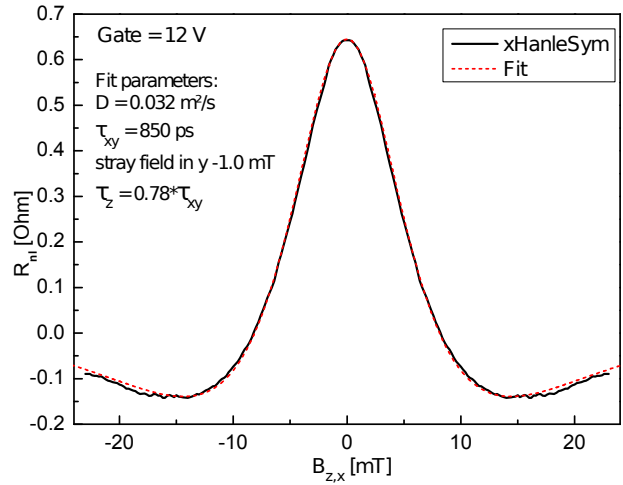

FIG. S12. Symmetrized xHanle data (black), with fit trace for $\zeta = 0.78$ which was judged a good fit.

which was done by manually adjusting the parameters and judging if the fit is good or not. This could be determined with an accuracy for $\zeta$ of $\pm 0.02$.

The fit for the symmetrized xHanle data is shown in Fig. S12, which is the fit we deemed to be optimal and gives $\zeta = 0.78$. This fit can be compared with Fig. S13, that shows a fit for $\zeta = 0.76$ in Fig. S13a) and a fit for $\zeta = 0.80$ in Fig. S13b). The fit for $\zeta = 0.76$ matches the shape very well for $B < \pm 10\,\text{mT}$, but it does not match the amplitude of the local minima at $B = \pm 15\,\text{mT}$. At these points the anisotropy would have the biggest influence, so matching these was a requirement for a good fit. The fit for $\zeta = 0.80$ matches the amplitude of the minima, but does not give a good fit for the shape. The fit for $\zeta = 0.78$ in Fig. S12 is the best compromise for matching shape and minima amplitude. As the shape of the Hanle curves was sometimes distorted due to drift or varying magnetic configuration (monitored by the difference between P and AP curves), we found that total accuracy of the fit, including both fitting precision and distortion, amounted to $\pm 0.04$.

The second source of uncertainty for the xHanle is the stray field in $y$. We fitted for a constant stray field of $1\,\text{mT}$ for xHanle and zHanle. If the stray field during xHanle measurement was different by $\pm 1\,\text{mT}$ to the stray field during zHanle measurement, that would result in an error for $\zeta$ of $\pm 0.02$. However, the occurrence of this error is unlikely, as the measurement sequence was:

Measure zHanle P, measure xHanle P, prepare AP orientation, measure zHanle AP, measure xHanle AP.

Here P signifies the parallel orientation of the electrodes and AP the antiparallel orientation. A change to the stray field in $y$ would most likely occur after the $y$ magnet was used, which in this sequence only happens during the

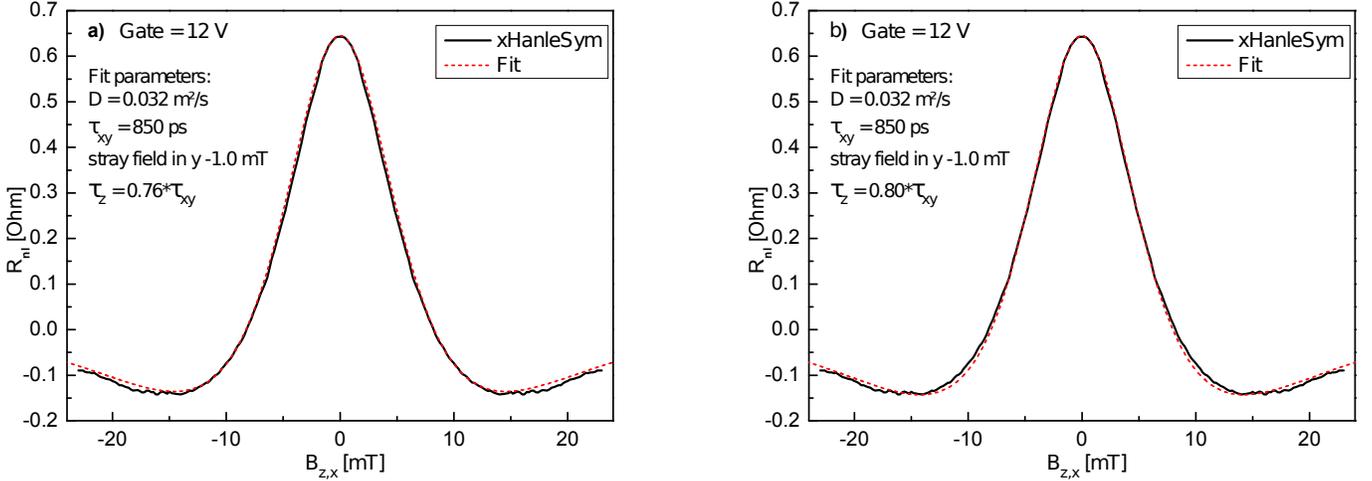

FIG. S13. Symmetrized xHanle data (black), with fit traces for a) $\zeta = 0.76$ and b) $\zeta = 0.80$, which were judged as inferior fits.

preparation of the AP orientation. Assuming this, the zHanle AP and xHanle AP measurement would have had a changed stray field. As these data are averaged with their P counterparts, the resulting average stray field for zHanle and xHanle would be the same.

For the total error of $\zeta$ obtained from the xHanle we then give $\pm 0.06$, which we estimate is the worst case scenario. As we think an error from the stray fields is unlikely, this number also includes other minor error source not considered here.

The uncertainty of the parameter $\zeta$ calculated from the oblique spin precession method is $\pm 0.15$ for an average $\zeta$ of 0.91 as displayed in Fig. 14 of the main text. The uncertainty is so high because there are several sources of uncertainty for the oblique spin precession.

We have five minor sources of error, which we estimate each to contribute an error of $\pm 0.01 \ldots 0.02$ to $\zeta$. The first is that we rotate the magnetic field by using the linear combination of two magnetic coils. Because of remanent magnetization in our magnets, we estimate that our magnetic field angle $\beta$ has some inaccuracies. B. Raes et al.[2] use a single magnet coil and instead rotate the sample, which is more accurate to determine $\beta$.

The second minor source of error is the fitting of the oblique spin precession data, which was done by hand and will have comparable inaccuracies as the xHanle fitting. However, the oblique spin precession data has to be normalized before it can be fitted, which we count as the third minor source of error.

A fourth source of error is the correction for the dynamic tilting of the electrodes that is expressed by $\gamma$. We calculate our $B_s$ from the data shown in Fig. S10b), and these data are quite noisy. The change in $\zeta$ because of $\gamma$ with -0.07 is not insignificant, so this needs to be considered as a potential source of minor uncertainty.

The fifth minor source of error is in the formula used for fitting, equation 7. More specifically, the formula contains the term $\sqrt{\frac{L^2}{\tau_{xy}D}}$. As displayed in Fig. 8 in the main text, we determined $\tau_{xy}$ with a certain amount of uncertainty. This uncertainty is then inherited by equation 7.

The biggest source of uncertainty in determining $\zeta$ from the oblique spin precession data comes from the tilted magnetization at the electrode interface that we describe with the tilt angle $\phi$. We used $\phi = 8.53°$ for the fit and this is responsible for a shift in $\zeta$ of -0.16. As detailed in section IV, estimating $\phi$ from the available data cannot be done with great accuracy. This is exemplified by the data in Fig. S9a), where a scaling factor of 0.3 ($\phi = 16.6°$) can also fit the data. We consider this high volatility unlikely and assume an uncertainty of $\pm 0.08$ because of the tilted magnetization. That would correspond to an error in $\phi$ of $\pm 5°$.

## VII. FINITE ELEMENT MODELING USING COMSOL

To model spin transport with arbitrary magnetic field direction and anisotropic spin relaxation times, we used the commercial finite element package COMSOL Multiphysics, Version 4.2a. This software includes predefined modules for many physical situations, such as the electric current module to model current flow in a sample of arbitrary geometry and given conductivity. However, there is no module to simulate spin transport. Therefore, we applied the coefficient form partial differential equation (PDE) module, which allows to enter a very general linear PDE of up to

second order by its coefficients. The general PDE predefined in COMSOL has the following form:

$$e_a \frac{\partial^2 \vec{u}}{\partial t^2} + d_a \frac{\partial \vec{u}}{\partial t} + \nabla \cdot (-c\nabla \vec{u} - \alpha \vec{u} + \gamma) + \beta \cdot \nabla \vec{u} + a\vec{u} = f \quad (8)$$

Here, we equate the COMSOL solution $\vec{u}$ to the desired spin density $\vec{s}$, and set the coefficients $e_a$, $\beta$ and $f$ to zero, as they do not appear in the diffusion equation for the spin density (Eq (1) of the main text).

The full spin drift-diffusion equation according to Fabian et al reads[3]:

$$\frac{\partial \vec{s}}{\partial t} - \nabla[\mu \vec{E}\vec{s} + D\nabla \vec{s}] = \vec{s} \times \vec{\omega_0} - \frac{\vec{s}}{\tau_s} \quad (9)$$

By multiplying out the precession term and relaxation term and assuming that the spin relaxation time can be different for each spatial direction we identify the last term in the COMSOL model with

$$a\vec{u} = -\vec{s} \times \vec{\omega_0} + \frac{\vec{s}}{\tau_s} = -\begin{pmatrix} -\frac{1}{\tau_x} & \omega_z & -\omega_y \\ -\omega_z & -\frac{1}{\tau_y} & \omega_x \\ \omega_y & -\omega_x & -\frac{1}{\tau_z} \end{pmatrix} \vec{s} \quad (10)$$

In the context of this work, $\tau_x = \tau_y$ equal to $\tau_{xy}$ in the main text.

Similarly, we identify the drift and diffusion terms in the spin equation with the corresponding COMSOL expressions

$$\alpha \vec{u} + c\nabla \vec{u} = \mu \vec{E}\vec{s} + D\nabla \vec{s} = \begin{pmatrix} \mu E_i & 0 & 0 \\ 0 & \mu E_i & 0 \\ 0 & 0 & \mu E_i \end{pmatrix} \vec{s} + \begin{pmatrix} D & 0 & 0 \\ 0 & D & 0 \\ 0 & 0 & D \end{pmatrix} \nabla \vec{s} \quad (11)$$

We do not consider drift due to the comparatively low currents involved and set the drift term to zero by assuming $\mu = 0$. The coefficient $\gamma$ in Eq. (8) can be used to model the spin Hall effect, which is not used here. All the coefficients we identified above can be typed into the COMSOL user interface either as numerical values or as variables which can be changed in a parametric sweep step.

For the sample boundaries, we apply the Zero Flux boundary condition in COMSOL, and for spin injection we set the Flux Source boundary condition at the position of the spin injecting electrode, with the injected spin current in the desired direction (e.g. along the stripes in $y$-direction, or including a magnetization tilt with a $z$-component ) proportional to the charge current normal to the interface. Both conditions are implemented with the equation:

$$-\vec{n} \cdot (-c\nabla \vec{u} - \alpha \vec{u} + \gamma) = g - q\vec{u} \quad (12)$$

where $\vec{n}$ is the vector normal to the boundary. The right hand side is set to zero for the Zero Flux condition, whereas the right hand side is set proportional to the electric current density for the Flux Source condition.

The geometry was always defined one-dimensional. Using a 2D model with the actual sample geometry resulted in virtually identical Hanle curves, so the simpler 1D model was used. The sample boundaries were extended to several spin flip lengths outside the region of interest to exclude any spin reflection in the confined geometry. Modeling the Pd electrodes at the correct distance as perfect spin sinks ($\vec{s}$ forced to zero) did not allow fitting the zHanle curves with reasonable $\tau_{xy}$ and the xHanle curves could not be fitted even for $\zeta = 0.5$. Modeling the Pd electrodes as spin reflecting instead resulted in essentially the same $\tau_{xy}$ and $\zeta$ as in the extended model. We therefore opted for the simpler extended geometry.

---



\end{document}